\documentclass[%
reprint,
superscriptaddress,
nofootinbib,
 amsmath,amssymb,
 aps,
prb,
]{revtex4-2}

\usepackage{graphicx}
\usepackage{dcolumn}
\usepackage{bm}
\usepackage{xcolor}

\def\muc{{\mu^*}} 
\newcommand{\ef}{\varepsilon_{\mathrm{F}}}
\newcommand{\nf}{f}
\newcommand{\nb}{n}
\newcommand{\NF}{N(\varepsilon_\text{F})}
\newcommand{\Ne}{N_\text{e}}
\newcommand{\muF}{\mu_{\rm{F}}}

\newcommand{\wj}{\omega_j}
\newcommand{\wjp}{\omega_{j'}}
\newcommand{\wc}{\omega_\text{c}}

\newcommand{\afkko}{\alpha^{2} F_{k,k'}(\Omega)}
\newcommand{\afo}{\alpha^{2} F(\Omega)}

\newcommand{\bk}{\mathbf k}

\begin{document}

\preprint{APS/123-QED}

\title{Fast Real-Axis Eliashberg Calculations:\\
Full-bandwidth solutions beyond the constant density of states approximation}

\affiliation{%
 Research Laboratory of Electronics, Massachusetts Institute of Technology \\
 50 Vassar Street, Cambridge, MA, USA, 02139-4307
}%

\affiliation{%
 Institute of Theoretical and Computational Physics, Graz University of Technology \\ 
 Petersgasse 16, 8010 Graz, Austria
}%

\author{Alejandro Simon$^\dagger$}
\email{alejansi@mit.edu}

\affiliation{%
 Research Laboratory of Electronics, Massachusetts Institute of Technology \\
 50 Vassar Street, Cambridge, MA, USA, 02139-4307
}%

\author{James Shi$^\dagger$}%

\affiliation{%
 Research Laboratory of Electronics, Massachusetts Institute of Technology \\
 50 Vassar Street, Cambridge, MA, USA, 02139-4307
}%

\author{Dominik Spath}
\affiliation{%
 Institute of Theoretical and Computational Physics, Graz University of Technology \\ 
 Petersgasse 16, 8010 Graz, Austria
}%

\author{Eva Kogler}
\affiliation{%
 Institute of Theoretical and Computational Physics, Graz University of Technology \\ 
 Petersgasse 16, 8010 Graz, Austria
}%

\author{Reed Foster}
\affiliation{%
 Research Laboratory of Electronics, Massachusetts Institute of Technology \\
 50 Vassar Street, Cambridge, MA, USA, 02139-4307
}%

\author{Emma Batson}
\affiliation{%
 Research Laboratory of Electronics, Massachusetts Institute of Technology \\
 50 Vassar Street, Cambridge, MA, USA, 02139-4307
}%

\author{Pedro N. Ferreira}
\affiliation{%
 Institute of Theoretical and Computational Physics, Graz University of Technology \\ 
 Petersgasse 16, 8010 Graz, Austria
}%

\author{Mihir Sahoo}
\affiliation{%
 Institute of Theoretical and Computational Physics, Graz University of Technology \\ 
 Petersgasse 16, 8010 Graz, Austria
}%

\author{Phillip D. Keathley}
\affiliation{%
 Research Laboratory of Electronics, Massachusetts Institute of Technology \\
 50 Vassar Street, Cambridge, MA, USA, 02139-4307
}%

\author{Warren E. Pickett}
\affiliation{%
 Department of Physics and Astronomy \\ 
 University of California Davis, Davis, California 95616, United States
}%

\author{Rohit Prasankumar}
\affiliation{%
 Deep Science Fund, Intellectual Ventures \\ 
 Intellectual Ventures, Bellevue, Washington, United States
}%

\author{Karl K. Berggren}
\affiliation{%
 Research Laboratory of Electronics, Massachusetts Institute of Technology \\
 50 Vassar Street, Cambridge, MA, USA, 02139-4307
}%

\author{Christoph Heil}
\affiliation{%
 Institute of Theoretical and Computational Physics, Graz University of Technology \\ 
Petersgasse 16, 8010 Graz, Austria
}%

\date{\today}

\begin{abstract}
Experimentally relevant signatures of superconductivity require access to real-frequency quantities, such as the spectral functions, optical response, and transport properties, yet Migdal–Eliashberg calculations are commonly performed on the imaginary axis and then analytically continued, a step that is numerically delicate and can obscure physically relevant spectral features. Here we present a practical route to solving the finite-temperature Migdal–Eliashberg equations directly on the real-frequency axis, while retaining the effects from the full-bandwidth electronic structure. Our formulation accounts for particle–hole asymmetry through an energy-dependent electronic density of states, avoiding the constant density of states approximation often used in real-axis calculations, and includes a static screened Coulomb contribution. We introduce an efficient numerical technique to solve the Migdal-Eliashberg integrals whose computational cost scales linearly with the real-frequency grid, making high-resolution, full-bandwidth real-axis calculations feasible and providing direct access to the interacting Green’s function and derived observables without analytic continuation.
As an illustration, we apply the method to H$_3$S, where a van-Hove singularity near the Fermi level produces strong particle–hole asymmetry. The full-bandwidth solution yields noticeably different spectra than the constant density of states approximation and brings the superconducting gap and lineshapes into closer agreement with experiment, highlighting when band-structure details are essential. Furthermore, the methods presented here open the door to time-dependent, nonequilibrium simulations within Eliashberg theory.
\end{abstract}

\def\thefootnote{$\dagger$}\footnotetext{These authors contributed equally to this work}
\maketitle

\section{introduction}

Most experimentally accessible signatures of conventional superconductors, such as tunneling spectra, optical conductivity, and transport properties, are intrinsically real-frequency quantities. The contribution to these quantities due to the electron-phonon interaction is treated within the framework of Migdal-Eliashberg theory, where one obtains the equilibrium properties and linear response of a material through the Green's function technique~\cite{migdal1958, eliashberg1960, ScalapinoSchrieffer, parks1969superconductivity, Pellegrini2024}. Yet, due to the singular nature of the self-energies, Migdal–Eliashberg calculations are still most often performed on the imaginary-frequency axis and then analytically continued to the real-axis~\cite{Vidberg1977, Marsiglio1988iterativeACON, Margine2013, PONCE2016116, fetter2012quantum, mahan, ALLEN19831, MARSIGLIO2020168102, isoME}. However, analytic continuation is an ill-conditioned procedure that can significantly amplify numerical errors. It is therefore a delicate step that may blur or distort fine spectral features and becomes increasingly challenging at low temperatures. Several studies have explored methods to improve the numerical procedure for performing the analytic continuation between the imaginary and real frequencies \cite{baker1961pade, kraberger2017maximum, vitali2010ab, PhysRevB.61.5147, khodachenko2024nevanlinna}, yet the ill-conditioned nature of the analytic continuation remains a fundamental limitation. Moreover, within the imaginary-axis formulation, nonequilibrium dynamics are cumbersome to deal with \cite{RevModPhys.58.323, 2011qftnbookR}. Recent work has explored an intermediate representation of the Green's function to resolve the problem of requiring finely spaced Matsubara frequencies at lower temperatures \cite{shinaoka2017compressing, chikano2019irbasis}; yet, this representation still does not recover the physical picture that a direct solution on the real-frequency axis would provide. At the same time, many existing real-axis treatments are computationally demanding and rely on a constant electronic density of states (cDOS) near the Fermi level \cite{holcomb1996finite}, effectively discarding full-bandwidth particle–hole asymmetry and band-structure features that can be important in real materials. 

These limitations, which constrain the computation of the full-bandwidth Green's function to the imaginary-frequency axis and require the ill-conditioned procedure of analytic continuation, often present problems in the theoretical study of superconductivity, where several quantities of interest, including the transport properties and nonequilibrium response of the material, can only be obtained from real-frequency-axis solutions to the Green's function. This often necessitates resorting to phenomenology to describe experimental data \cite{ghosh2019strong}. 

\begin{figure*}
    \centering
    \includegraphics[width=\linewidth]{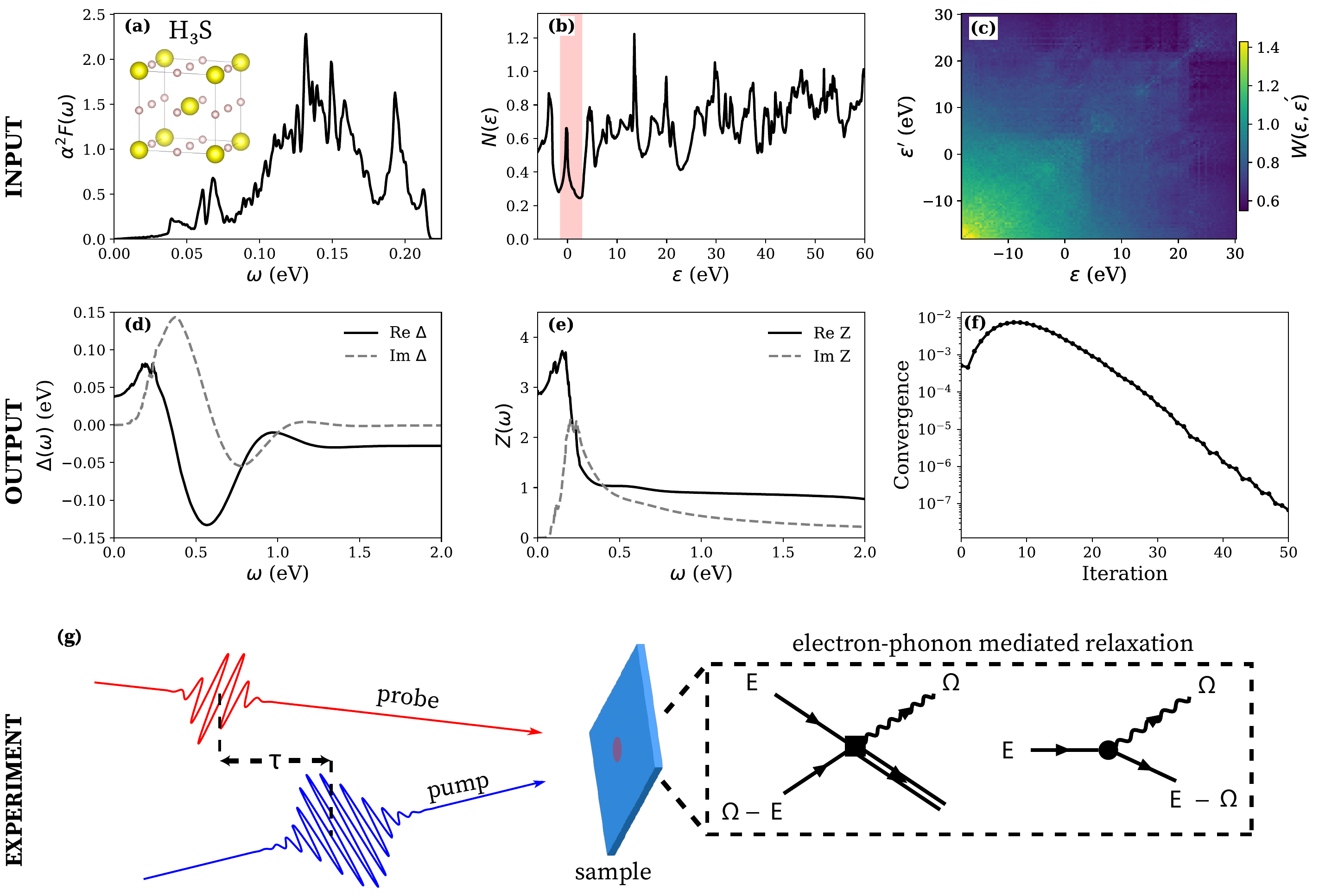}
    \caption{The superconducting and electronic properties of H$_3$S derived from its crystal structure. (a) The Eliashberg spectral function $\alpha^2F(\omega)$ obtained from density functional perturbation theory. (b) The electronic density of states $N(\varepsilon)$ for H$_3$S computed using density functional theory. The van-Hove singularity near the Fermi level ($\varepsilon = 0$) is highlighted in red. (c) The effective screened Coulomb potential $W(\varepsilon,\varepsilon')$ as a function of electron energy $\varepsilon$ calculated from the GW approximation. These quantities are used as input for numerical solutions to the Migdal-Eliashberg equations. The corresponding superconducting gap edge (d) $\Delta(\omega)$, (e) renormalization parameter $Z(\omega)$, and (f) root-mean square deviation between iterations of $\Delta(\omega)$ (convergence) in the constant density of states and $\mu^*$ approximation at $T=1\,$mK. (g) With access to the real-axis solutions, one can compute the transport properties of a material and use it to model the response in nonequilibrium perturbations, such as in pump-probe experiments.}
    \label{fig:figure1}
\end{figure*}

In this work, we formulate and efficiently solve the finite-temperature Migdal–Eliashberg equations directly on the real-frequency axis while retaining an energy-dependent electronic density of states, enabling direct computation of the interacting Green’s function and real-frequency observables without analytic continuation. In the following, we begin by deriving a form for the real-axis Migdal-Eliashberg equations that is amenable to numerical computation. We then provide numerical techniques to efficiently evaluate the real-axis Migdal-Eliashberg equations. In contrast to previous implementations of real-frequency axis solutions that remain computationally expensive to obtain \cite{holcomb1996finite, Marsiglio1988iterativeACON, Margine2013}, we present a numerical technique that has a computational complexity linear in the number of points used to sample the integrals in the Migdal-Eliashberg equations. We also generalize our approach to incorporate the particle-hole asymmetry introduced by considering a variable electronic density of states, similarly to the full bandwidth approach on the imaginary-frequency axis in Ref. \cite{lucrezi_full-bandwidth_2024}. Using this method, we then examine the effect of the van-Hove singularity (vHS) on the superconducting properties and excitation spectra in H$_3$S. 

The resulting framework enables detailed, first-principles studies of superconducting materials and their dynamical response, as demonstrated in Ref. \cite{FIXME-THIS-PAPER}. Figure \ref{fig:figure1} summarizes the workflow: subfigures (a)–(c) provide the \textit{ab initio} inputs, computed via density functional perturbation theory \cite{Pellegrini2024, isoME}, to a real-frequency Migdal–Eliashberg solver. Subfigures (d) and (e) show representative outputs for the superconducting gap $\Delta(\omega)$ and the renormalization function $Z(\omega)$ within the cDOS approximation, while (f) illustrates the rapid convergence of our fixed-point iteration through the decreasing RMS deviation of $\Delta(\omega)$ between iterations. With fast access to full-bandwidth solutions for $\Delta(\omega)$ and $Z(\omega)$, we can compute transport and optical properties within Eliashberg theory, including nonequilibrium responses such as those probed in pump–probe experiments on superconducting films (g) \cite{FIXME-THIS-PAPER}.

\section{Full-Bandwidth Real-Axis Eliashberg Equations}

We start by laying out the finite-temperature Migdal–Eliashberg framework in a form suited for direct real-frequency calculations. Working in the anisotropic approximation, we express the self-energy in Nambu space as:~\cite{migdal1958, eliashberg1960, Nambu, ALLEN19831}
\begin{equation}
\label{eq:SelfEnergyGeneral}
\begin{aligned}
    \Sigma(k,i\wj) = & -\frac{1}{\beta}\sum_{k',j'}\tau_3 G(k,i\wjp) \tau_3 \Bigg[W_{k,k'}(i\wj-i\wjp) \\ & + \sum_\lambda |g_{kk'\lambda}|^2 D_\lambda(k-k',i\wj-i\wjp)\Bigg],
\end{aligned}
\end{equation}
where $\beta= 1/ k_{\mathrm{B}} T$ is the inverse temperature, $k=(\bk,n)$ is a combined momentum and band index, $\lambda$ the phonon mode index, $\tau_i$ is the $i$-th Pauli matrix, $G$ and $D_\lambda$ are the electron and phonon thermodynamic Green's functions respectively, $W_{k,k'}$ describes the Coulomb interaction between pairs of electrons, and $g_{kk'\lambda}$ describes the electron-phonon coupling. Inserting the spectral representation of the electron $A(k,\omega)$ and phonon $B(k,\Omega)$ Green's functions~\cite{ALLEN19831, marsiglio2001}, we get
\begin{equation}
\begin{aligned}
\Sigma(k,i\wj) = & -\frac{1}{\beta}\sum_{k',j'} \int d\omega' \frac{A(k',\omega')}{i\wjp-\omega} \Bigg[W_{k,k'} \\&+ \frac{1}{\NF} \int_{-\infty}^\infty d\Omega \frac{\afkko}{i\wj-i\wjp-\Omega} \Bigg]~,
\end{aligned}
\end{equation}
where the Eliashberg spectral function has been introduced~\cite{ALLEN19831}:
\begin{equation}
    \afkko = \NF \sum_\lambda  |g_{kk'\lambda}|^2 B_\lambda(k-k',\Omega)~,
\end{equation}
and $N(\ef)$ is the single-spin electronic density of states at the Fermi level $\ef$.
Performing the sum over Matsubara frequencies and analytically continuing $i\omega_j \mapsto \omega + i0^+$, we obtain
\begin{widetext}
\begin{align}
    \Sigma(\varepsilon, \omega) = -\frac{1}{\pi} &\int_{-\infty}^{\infty} d\omega'  \int_{-\infty}^{\infty} d\varepsilon' N(\varepsilon') \Bigg\{ \Im[\tau_3G(\varepsilon',\omega')\tau_3]\frac{K(\omega,\omega')}{\NF}
        - \frac{1}{2} \Im[\tau_3G^{\text{od}}(\varepsilon',\omega')\tau_3]W(\varepsilon,\varepsilon')\big[2\nf(\omega')-1\big] \Bigg\}, \nonumber
\end{align}
\end{widetext}
where \\\\\\
\begin{align} 
    K(\omega, \omega') = & \int_{0}^{\infty} d\Omega \afo \\ & \times \bigg[\frac{\nb(\Omega)+1-\nf(\omega')}{\omega-\Omega-\omega'+i0^+}+\frac{\nb(\Omega)+\nf(\omega')}{\omega+\Omega-\omega'+i0^+} \bigg] \nonumber
\end{align}
is the integral kernel, and the Coulomb contribution to the self-energy contains only the off-diagonal elements of $G^{\text{od}}(k,i\wj)$ since the Coulomb interaction is already contained in the band structure of the normal state~\cite{ALLEN19831}. Here, we  have also substituted in the definition for the spectral function 
\begin{equation}
    A(k,\omega) = -\frac{1}{\pi}\Im \{\tau_3 G(k,\omega+i0^+)\tau_3\}
\end{equation} 
and replaced the sum over $k'$ with a weighted integral over the electronic density of states $N(\varepsilon')$. The self-energy can be cast in a more familiar form in terms of its Pauli-matrix decomposition~\cite{ScalapinoSchrieffer}:
\begin{equation}
\Sigma(\varepsilon, \omega) = [1-Z(\omega)]\omega \tau_0 + \phi(\varepsilon, \omega) \tau_1 + \chi(\omega) \tau_3~,
\end{equation}
where we have neglected the $\varepsilon$ dependence of $Z$ and $\chi$~\cite{ALLEN19831}.
Next, through the Dyson equation $G^{-1}(\varepsilon,\omega) = G_0^{-1}(\varepsilon,\omega) - \Sigma(\varepsilon,\omega)$ with the non-interacting Green's function given by $G_0(\varepsilon,\omega) = 1/[\omega - (\varepsilon - \mu_F) + i0^+]$, we obtain a decomposition of the interacting Green's function~\cite{ALLEN19831, PONCE2016116}:
\begin{equation}
G(\varepsilon,\omega) = \frac{\omega Z(\omega)\tau_0 + [\varepsilon - \mu_F + \chi(\omega)]\tau_3 + \phi(\varepsilon,\omega)\tau_1}{\Theta(\varepsilon,\omega)}
\end{equation}
where the determinant is given by 
\begin{equation*}
\Theta(\varepsilon,\omega) = \omega^2 Z^2(\omega) - [\varepsilon - \mu_F + \chi(\omega)]^2 - \phi(\varepsilon,\omega)^2.
\end{equation*}
Substituting this form into the self-energy expression yields the full isotropic Migdal-Eliashberg equations on the real-frequency axis:
\begin{widetext}
\begin{subequations}
    \begin{align}
        \label{eq:realvDOSW_Z_main}
        Z(\omega) &= 1 + \frac{1}{\omega\pi\NF} \int_{-\infty}^\infty d\omega' \, K(\omega, \omega') \int_{-\infty}^\infty d\varepsilon' \, N(\varepsilon') \, 
        \Im \left[\frac{Z(\omega')\omega'}{\Theta(\varepsilon',\omega')}\right] \\
        \label{eq:realvDOSW_chi_main}
        \chi(\omega) &= - \frac{1}{\pi\NF} \int_{-\infty}^\infty d\omega' \, K(\omega, \omega') \int_{-\infty}^\infty d\varepsilon' \, N(\varepsilon') \,
        \Im \left[\frac{\varepsilon'-\mu_{\rm{F}}+\chi(\omega')}{\Theta(\varepsilon',\omega')}\right] \\
        \label{eq:realvDOSW_phi_main}
        \phi(\varepsilon,\omega) &= \frac{1}{\pi\NF} \int_{-\infty}^\infty d\omega' \int_{-\infty}^\infty d\varepsilon' \, N(\varepsilon') \,
        \Im \left[\frac{\phi(\varepsilon',\omega')}{\Theta(\varepsilon',\omega')}\right] 
        \left\{ K(\omega, \omega') - \frac{1}{2}\NF W(\varepsilon, \varepsilon')[2f(\omega') - 1] \right\} \\
        \label{eq:num_e_main}
        n_e &= \int_{-\infty}^{\infty} d\varepsilon' \, N(\varepsilon') \left[ 1 - \frac{1}{\pi} \int_{-\infty}^{\infty} d\omega' \, \tanh\left( \frac{\omega'}{2 k_B T} \right)
        \Im \left\{ \frac{\varepsilon' - \muF + \chi(\omega')}{\Theta(\varepsilon',\omega')} \right\} \right]
    \end{align}
\end{subequations}
\end{widetext}
where $Z(\omega)$ is the wavefunction renormalization parameter, $\phi(\varepsilon,\omega)$ is the superconducting order parameter, and $\chi(\omega)$ is the effective chemical potential shift of the electrons. Further details regarding the derivation and details on the standard approximations typically used to reduce Eqs. \eqref{eq:realvDOSW_Z_main}, \eqref{eq:realvDOSW_chi_main}, \eqref{eq:realvDOSW_phi_main}, and \eqref{eq:num_e_main} are provided in the Supplemental Information (SI). Here, $N(\varepsilon)$ is the electronic density of states, $\ef$ is the Fermi energy, $f(\omega)$ and $n(\Omega)$ are the quasiparticle and phonon distributions, respectively. At thermal equilibrium, $f(\omega)$ and $n(\Omega)$ are the usual Fermi and Bose distributions. Henceforth, we refer to the variable density of states (vDOS) as the full set of equations defined by Eqs. \eqref{eq:realvDOSW_Z_main}, \eqref{eq:realvDOSW_chi_main}, \eqref{eq:realvDOSW_phi_main}, and \eqref{eq:num_e_main}, in which both the electronic density of states $N(\varepsilon)$ and the screened Coulomb interaction $W(\varepsilon, \varepsilon')$ retain their energy dependence~\cite{lucrezi_full-bandwidth_2024, Sano-2016, sanna-2018, Wang2020, pellegrini_eliashberg_2022, Pellegrini2024, Davydov_2020, isoME}. The constant density of states approximation (cDOS) refers to the approximation where $N(\varepsilon) \rightarrow N(\ef)$ is taken to be constant and the Coulomb contribution to the self-energy is reduced to the Morel-Anderson Coulomb pseudo-potential $\mu^*$ \cite{PhysRev.125.1263}.

\begin{figure*}
    \centering
    \includegraphics[width=\linewidth]{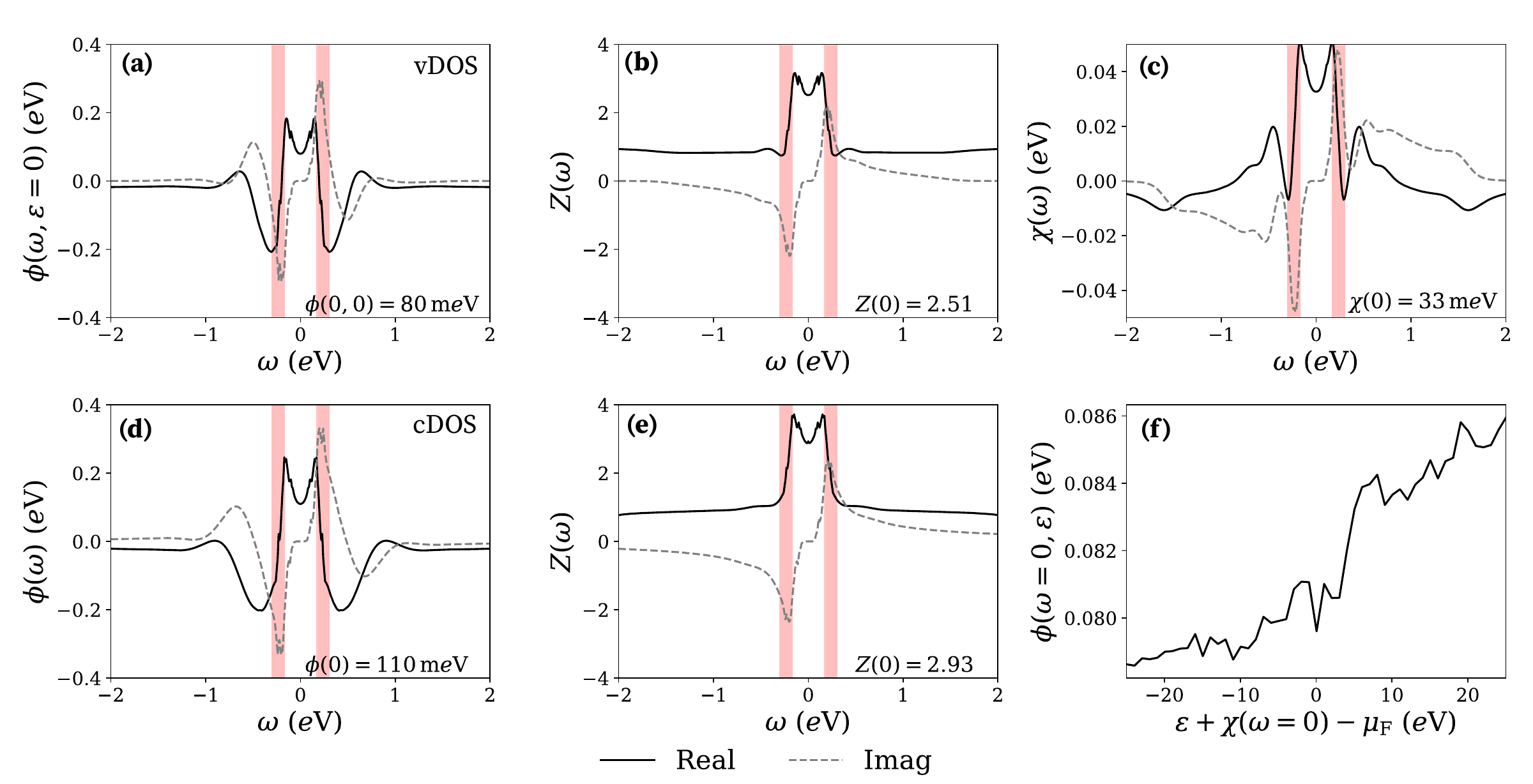}
    \caption{Solutions to the Migdal-Eliashberg equations for H$_3$S at $T=1\,$mK. The top panel corresponds to solutions considering the full variable electronic density of states (vDOS) for (a) the superconducting order parameter at the Fermi level $\phi(\omega,\varepsilon_k=0)$, (b) the wavefunction renormalization parameter $Z(\omega)$, (c) the shift in the chemical potential due to superconducting correlations $\chi(\omega)$, and (f) superconducting order parameter at the gap-edge as a function of $\varepsilon$. The bottom panel corresponds to the solution to the Migdal-Eliashberg equations assuming a constant electronic density of states at the Fermi level (cDOS) for (d) $\phi(\omega)$ and (e) $Z(\omega)$. In each panel, the real part corresponds to the solid black line, and the grey dashed line indicates the imaginary part. The solid red bars indicate the location of the van-Hove singularity in H$_3$S. The vDOS solution provides results closer in agreement with experiment.}
    \label{fig:figure2}
\end{figure*}

\section{Numerical approach}

To determine $G(\varepsilon,\omega)$, Eqs. \eqref{eq:realvDOSW_Z_main}, \eqref{eq:realvDOSW_chi_main}, \eqref{eq:realvDOSW_phi_main}, and \eqref{eq:num_e_main} must be solved together. To this end, we may take advantage of several features of the equations. Firstly, the integral kernel $K(\omega, \omega')$ contains the entire temperature dependence of the Migdal-Eliashberg equations. Thus, as pointed out by Holcomb \cite{holcomb1996finite}, it can be computed once for a given temperature and reused in the computation of each equation. However, precomputation of these kernels on a square grid of $(\omega,\omega')$ remains inefficient, as the runtime then scales quadratically with the density of the grid on the $\omega$ axis. Our approach to address this begins by noting that the real part of the integral kernel can be cast in the form
\begin{equation}
\label{eq:kernel-split}
\begin{aligned}
    \Re\{K(\omega, \omega')\} =& [-I_1(\omega -\omega') - (1-f(\omega'))I_2(\omega-\omega')] \\
    &+ [I_1(\omega'-\omega) + f(\omega')I_2(\omega'-\omega)]
\end{aligned}
\end{equation}
where 
\begin{equation*}
\begin{aligned}
    &I_1(x) = \mathcal{P} \int_{-\infty}^\infty d\Omega \, \alpha^2F(\Omega) \frac{n(\Omega)}{\Omega - x}\\
    &I_2(x) = \mathcal{P} \int_{-\infty}^\infty d\Omega \, \alpha^2F(\Omega) \frac{1}{\Omega - x}~,
\end{aligned}
\end{equation*} where $\mathcal{P}$ denotes the principal value. To compute $\Re\{K(\omega,\omega')\}$, we thus only need to evaluate $I_1(x)$ and $I_2(x)$ for all values of $\omega-\omega'$ and combine them using Eq. \eqref{eq:kernel-split}. If a linear grid of $(\omega,\omega')$ is used, then the number of possible values of $\omega-\omega'$ on the grid scales linearly with the number of points sampled on the $\omega$ axis. The imaginary component of the integral kernel $\Im\{K(\omega,\omega')\}$ is computed analytically, as the corresponding integrals reduce to an evaluation over a delta function. 

The linear scaling enabled by our approach allows for the use of a much denser grid while also providing a very efficient computation. In the SI, we further quantify the runtime improvement with our approach. This efficiency is crucial for time-dependent analysis because $K(\omega,\omega')$ must be recomputed at each time step as $f(\omega)$ and $n(\omega)$ may vary in nonequilibrium case. In Ref. \cite{FIXME-THIS-PAPER}, we combine the methods proposed in Refs. \cite{simon2025abinitiomodelingnonequilibrium, simon2025ab} with the fast algorithms developed here to perform non-equilibrium modeling within the framework of Migdal-Eliashberg theory.

\begin{figure*}
    \centering
    \includegraphics[width=\linewidth]{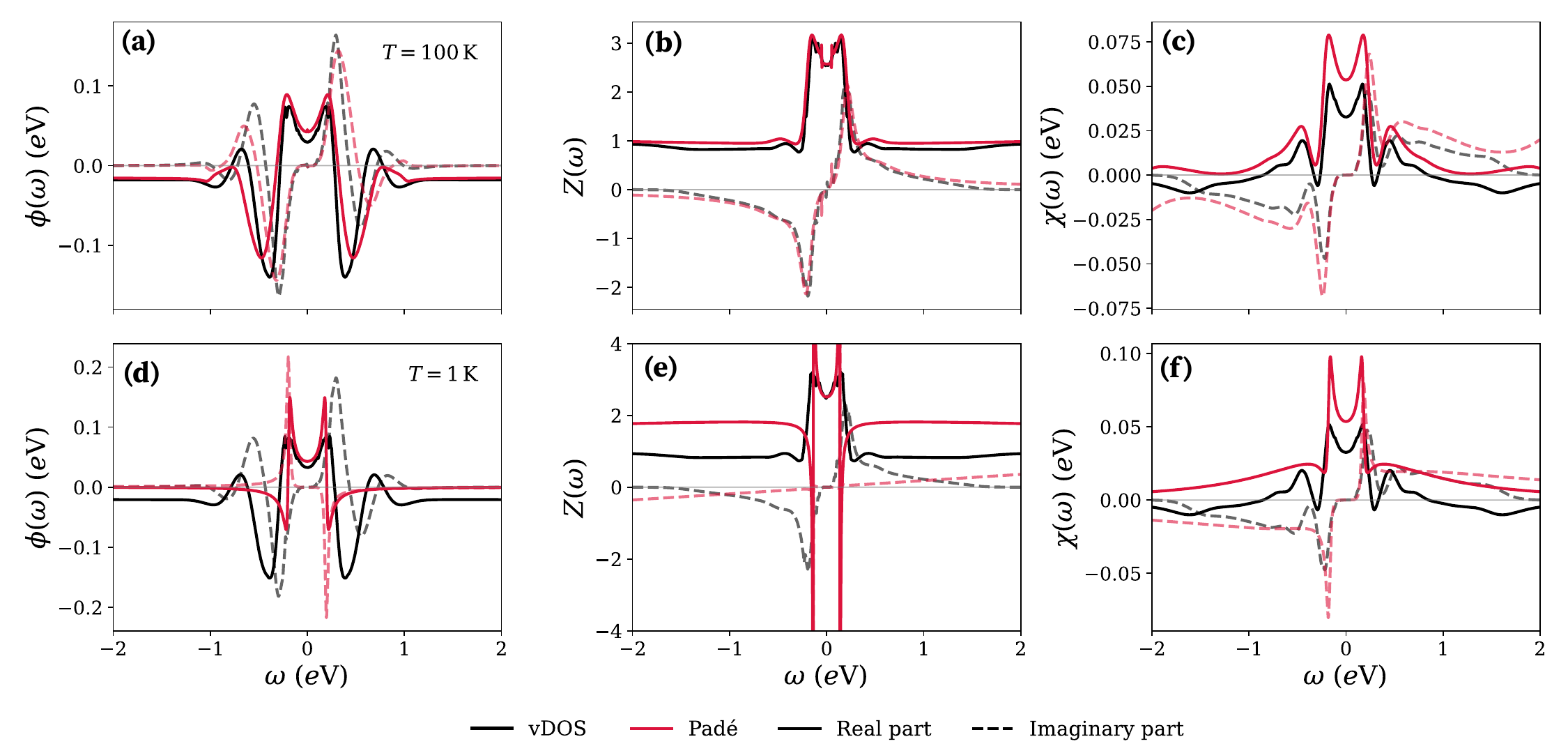}
    \caption{A comparison between solutions of the Migdal-Eliashberg equations for (a,d) $\phi(\omega)$, (b,e) $Z(\omega)$, and (c,f) $\chi(\omega)$ at (a-c) $T=100\,$K and (d-f) $T=1\,$K obtained with the direct real-frequency axis solver shown in black and the analytic continuation done with the Padé approximation shown in red. Numerical instabilities are clearly present in the analytically continued solutions, with large unphysical gradients notably appearing in (a), (b), (d), and (e).}
    \label{fig:main-Pade}
\end{figure*}

\begin{figure*}
    \centering
    \includegraphics[width=\linewidth]{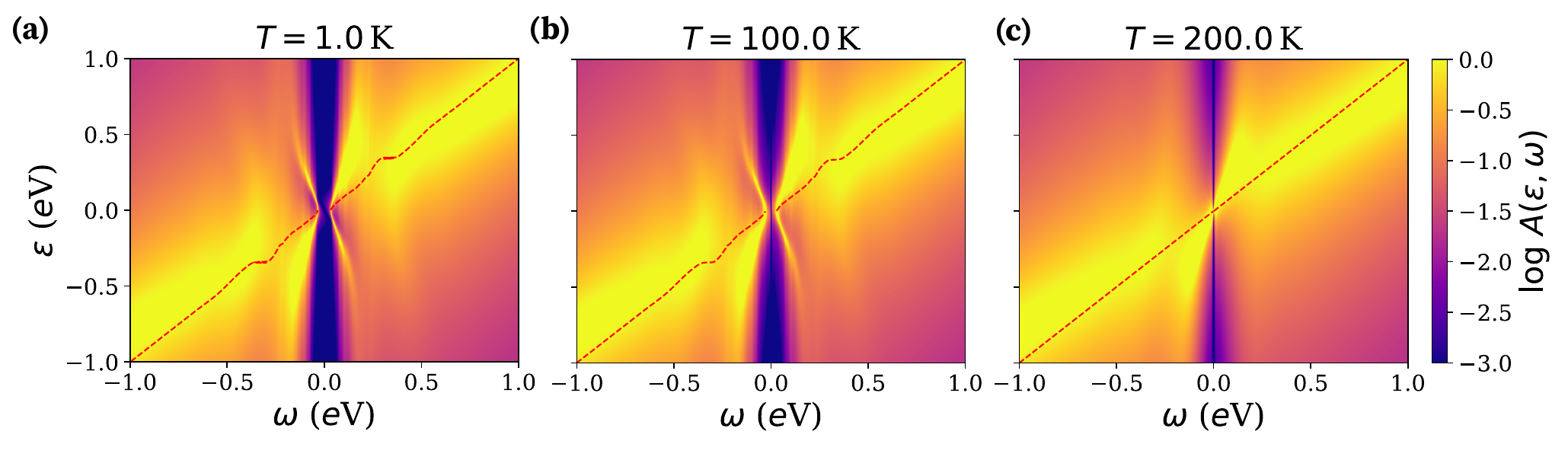}
    \caption{(a) The spectral function $A(\varepsilon,\omega)$ for H$_3$S at (1) $T=1\,$K, (b) $100\,$K, and (c) $200\,$K. The dashed red line indicates the Bogoliubov dispersion relation. Bright yellow regions indicate high spectral weight, whereas dark purple regions indicate low or zero spectral weight.}
    \label{fig:figure3}
\end{figure*}

In addition to the integral kernels, we must evaluate to high precision integrals of the spectral functions, which are of the form
\begin{equation}
    \int_{-\infty}^\infty d\varepsilon\, N(\varepsilon) \Im\left[\frac{g(\varepsilon, \omega)}{\Theta(\varepsilon, \omega)} \right],
\end{equation} where $g(\varepsilon,\omega)$ is defined in Eqs. \eqref{eq:realvDOSW_Z_main}, \eqref{eq:realvDOSW_chi_main}, and \eqref{eq:realvDOSW_phi_main}.
In the cDOS approximation, these integrals can be evaluated analytically using the residue theorem. However, when $N(\varepsilon)$ is allowed to vary arbitrarily, these integrals must be solved numerically except in specific cases. The resulting integrals are challenging to perform numerically due to sharp peaks near the poles of the Green's function.

To address these challenges, we use the fact that the spectral factors multiplying $N(\varepsilon)$ in the Migdal-Eliashberg equations have an analytic antiderivative, which is derived in the SI. In fact, any polynomial of $\varepsilon$ multiplied by this factor also has an analytic antiderivative. Thus, we can perform the integral by breaking it up into a sum of integrals along subintervals of width $d\varepsilon'$. For each subinterval, we interpolate $N(\varepsilon')$ linearly and perform the integral over $\varepsilon'$ analytically. So long as $N(\varepsilon')$ varies slowly with respect to the step size $d\varepsilon'$, this approach provides a low-error and efficient method to compute the spectral integrals. In the limit of infinitely dense sampling of $N(\varepsilon')$, this approximation approaches the exact integral value. In the case where $W(\varepsilon, \varepsilon')$ is not constant and $\phi(\varepsilon, \omega)$ has non-trivial $\varepsilon$-dependence, we also fit a piecewise-linear approximation to $\phi(\varepsilon,\omega)$ along the $\varepsilon$ axis. 

This numerical approach outlined here provides an efficient method to evaluate the integrals in the Migdal-Eliashberg equations to high precision on the real-frequency axis at finite temperature. We find that combining our numerical approach with a simple fixed-point iteration with a stopping criteria of less than a $0.01\%$ variation in $\phi(0,0)$ yields convergent solutions for $Z(\omega), \chi(\omega), \phi(\varepsilon, \omega)$, and $\mu_F$ that are generally within 1\% of the values computed on the imaginary-frequency axis with the IsoME package \cite{isoME}. When computed in the cDOS case, these highly converged results take only on the order of milliseconds to compute on typical modern laptop hardware, and this scales to minutes in the vDOS case. Because of the speed of the cDOS solutions, we use the cDOS solution as an initial guess for the vDOS fixed-point solve, which further speeds up computation. Implementing even more advanced methods, in particular for evaluating integrals over the spectral functions, will further improve stability and speed, and is the focus of current work. 

\section{Results and Discussion}

\begin{figure*}
    \centering
    \includegraphics[width=\linewidth]{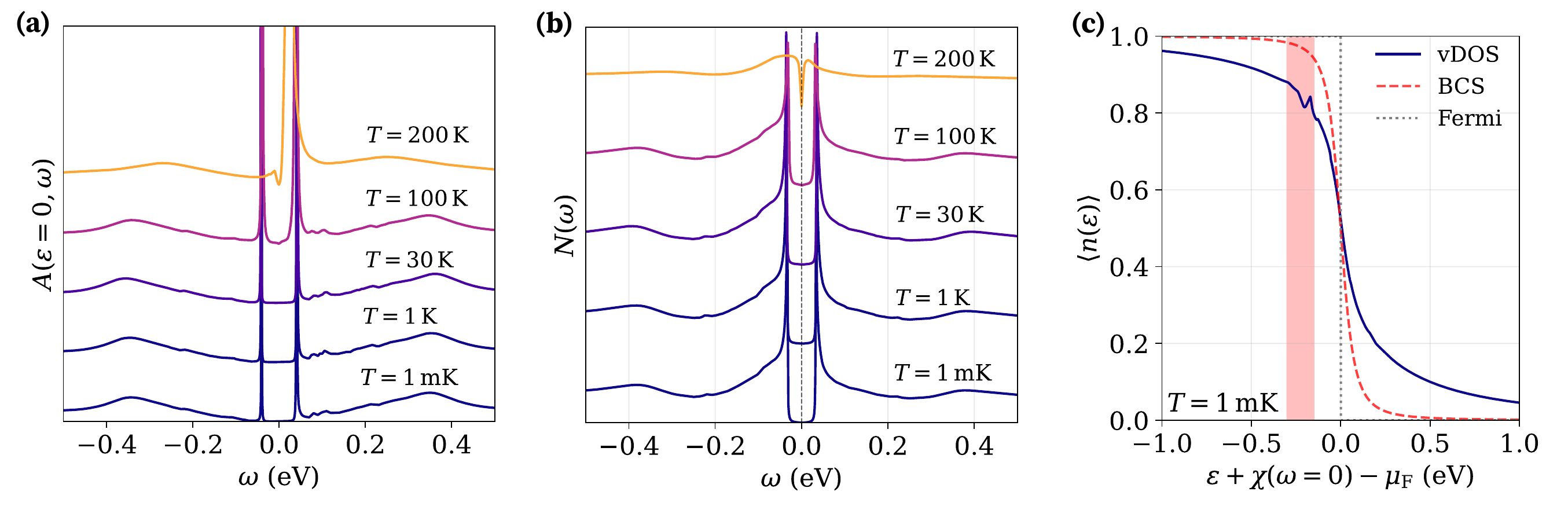}
    \caption{(a) The spectral function $A(\varepsilon=0,\omega)$ at $T=1\,$mK, $1\,$K, $30\,$K, $100\,$K, and $T=200\,$K. (b) The corresponding quasiparticle density of states $N(\omega)$ is computed from $A(\varepsilon, \omega)$. In (a) and (b), each temperature is offset on the y-axis for clarity. (c) The interacting quasiparticle occupancy $\langle n(\varepsilon) \rangle$ at $T=1\,$mK plotted alongside the BCS and non-interacting (Fermi-Dirac) distributions for comparison. The solid red bar indicates the location of the van-Hove singularity in H$_3$S.}
    \label{fig:figure4}
\end{figure*}

The Migdal-Eliashberg equations require material-specific inputs for $\alpha^2F(\omega)$, $N(\varepsilon)$, and $W(\varepsilon, \varepsilon')$. In Figure \ref{fig:figure1}a, b, and c, we present $\alpha^2F(\Omega)$, $N(\varepsilon)$, and $W(\varepsilon, \varepsilon')$ respectively, computed from density functional theory (DFT)~\cite{QE-2017}, density functional perturbation theory (DFPT), and $GW$ calculations~\cite{DESLIPPE20121269} for H$_3$S at 200 GPa. The solution to the Migdal-Eliashberg equations for H$_3$S at $T=1\,$mK in the cDOS and $\mu^*$ approximation are shown in Figure \ref{fig:figure1}d,e. The corresponding convergence, defined as the root mean square variation of $\Delta(\omega) = \phi(\omega)/Z(\omega)$ between iterations, is shown in Figure \ref{fig:figure1}f.

H$_3$S exhibits several features that make it an ideal testbed for our methods. As is common for hydrides, the light hydrogen atoms produce strong electron-phonon coupling and a large maximum phonon frequency $\omega_{\mathrm{D}}$ (Figure \ref{fig:figure1}a), yielding an experimental $T_{\mathrm{c}} \approx 200\,$K and superconducting gap $2\Delta(\omega=0) \approx 60\,$meV at 200 GPa \cite{drozdov2015conventional, du2025superconducting}. More importantly for this work, H$_3$S features a prominent van Hove singularity near the Fermi level (highlighted in red in Figure \ref{fig:figure1}b). This vHS induces strong particle-hole asymmetry that significantly affects the superconducting properties and manifests in macroscopic tunneling measurements \cite{quan2016van, ghosh2019strong, du2025superconducting}, making H$_3$S an excellent case for demonstrating the importance of the variable density of states treatment.

Using the values obtained from DFT and DFPT for H$_3$S shown in Figure \ref{fig:figure1}, we solved the full real-frequency axis Migdal-Eliashberg equations with the variable $N(\varepsilon)$. The solutions for $\phi(\omega,\varepsilon=0)$, $Z(\omega)$, $\chi(\omega)$, and $\phi(\omega=0,\varepsilon)$ at $T=1\,$mK are displayed in Figure \ref{fig:figure2}a,b,c,d, respectively. For comparison, the cDOS approximation for $\phi(\omega)$ and $Z(\omega)$ is plotted in Figure \ref{fig:figure2}e,f. Notably, the real-frequency dependence obeys the expected symmetry relations as required to preserve the causality of the Green's function and the Kramers-Kronig relation. A derivation of the expected symmetry relations for $Z(\omega)$, $\phi(\omega,\varepsilon)$, and $\chi(\omega)$ can be found in the SI. One can also leverage these known symmetry relations to further reduce the total number of points needed to sample the Migdal-Eliashberg integrals. The low-temperature zero frequency value of the superconducting gap $2\Delta(\omega=0,\varepsilon=0)=2\phi(\omega=0,\varepsilon=0)/Z(\omega=0)$ for the vDOS calculation is $\sim$60$\,$meV, which matches the experimental value \cite{du2025superconducting}. In contrast, the cDOS and $\mu^*$ approximation predicts 75$\,$meV. Both the cDOS and vDOS zero-frequency results obtained on the real axis are nearly identical to the corresponding values computed using conventional imaginary-axis methods \cite{isoME}. 

We also confirmed that our real-frequency axis solutions agree qualitatively with the analytic continuation of the imaginary-axis solutions obtained with the Padé approximation. This comparison is shown in Figure \ref{fig:main-Pade} and in more detail in the SI. Here, the imaginary frequency solutions were obtained with the IsoME package \cite{isoME}, and we used a smearing of $1\,\mathrm{m}e\mathrm{V}$ in the continuation procedure to ensure stability of the continued solution, while keeping all numerical parameters consistent between solutions. As we have previously pointed out, the ill-conditioned nature of the analytic continuation results in significant numerical instabilities, which are particularly evident in the large peaks in the analytic continuation for $Z(\omega)$ at $T=1\,\mathrm{K}$ in Figure \ref{fig:main-Pade}. Additionally, the fine structure in solutions for $\phi(\omega)$ is obscured in the analytically continued solutions. On the other hand, the direct real-frequency solutions we produce with our approach do not possess these instabilities and retain the fine structure introduced by $\alpha^2F(\omega)$ and $N(\varepsilon)$ over the entire temperature range.

The closer agreement of the vDOS solution with experiment confirms the importance of treating the electronic structure properly in H$_3$S \cite{ghosh2019strong}. Our real-frequency-axis approach allows us to quantify this effect directly: the vHS causes $\chi(\omega)$ to vary rapidly and substantially within the relevant energy range, manifesting as strong particle-hole asymmetry in the spectral functions, quasiparticle density of states, and occupancies shown in Figures \ref{fig:figure3} and \ref{fig:figure4}.

In Figure \ref{fig:figure3}, we show the resulting spectral function $A(\varepsilon,\omega)=-1/\pi \, \Im \, G(\varepsilon,\omega)$ at $T = 1\,$K, $100\,$K, and $200\,$K. At frequencies $\omega < \Delta(\omega=0)$, as expected, a clear gap in the excitation spectra is present, corresponding to the superconducting gap and the necessary $2\Delta(\omega=0)$ required to break a Cooper pair. Moreover, a ``butterfly'' shape near $\omega=\varepsilon_k=0$ corresponding to Bogoliubov quasiparticle dispersion relation \begin{equation}
    \omega = \pm E_k = \pm \sqrt{\varepsilon_k^2 + \Delta^2(\omega)},
\end{equation} is visible in $A(\varepsilon_k,\omega)$. The exact dispersion relation is plotted on top of the spectral functions in Figure \ref{fig:figure3} for reference. At large $\varepsilon_k \gg \Delta(\omega=0)$, the superconducting correlations become weak and the excitation spectrum approaches that of a normal metal. In the range of frequencies below $\omega_{\mathrm{D}}$, features of $\alpha^2F(\omega)$ lead to fine structure, which become particularly visible in the quasiparticle density of states $N(\omega)$. As $T \rightarrow T_{\mathrm{c}}$, the superconducting gap closes $\Delta(\omega=0)\rightarrow0$, and the Bogoliubov quasiparticle dispersion approaches that of a normal metal with $\omega = \pm \varepsilon_k$.  

Figure \ref{fig:figure4}a shows the temperature dependence of the spectral function $A(\varepsilon_k,\omega)$ for $\varepsilon_k=0$ from $T=1\,$mK to $200\,$K. A clear particle-hole asymmetry is observed when comparing the electron ($\omega>0$) and hole ($\omega<0$) branches of $A(\varepsilon_k=0,\omega)$. Integrating the $N(\varepsilon_k)$-weighted spectral function with respect to $\varepsilon_k$ results in the quasiparticle density of states shown in Figure \ref{fig:figure4}b. Here, $N(\omega)$ also reflects the strong particle-hole asymmetry of H$_3$S with a clear asymmetry across $\omega=0$. Moreover, features of $\alpha^2F(\omega)$ appear in the range $\omega < \omega_{\mathrm{D}}$ due to the strong-coupling in H$_3$S. Finally, integration of the spectral function with respect to $\omega$ from $(-\infty, 0]$ gives the ensemble-averaged quasiparticle occupancy function $\langle n(\varepsilon_k) \rangle$, which is also equal to the square of the quasiparticle amplitude $|v_k|^2$. The interacting $\langle n(\varepsilon_k) \rangle$ is shown alongside the BCS result for $\langle n(\varepsilon) \rangle$ in Figure \ref{fig:figure4}c. For clarity in Figure \ref{fig:figure4}c, the BCS distribution is offset on the $\varepsilon_k$-axis by $\chi(\omega)-\mu_{\mathrm{F}}$, so that the zero energy region of both distributions line up. The strong electron-phonon interaction leads to a shift in the chemical potential and a small kink in the $\varepsilon_k < 0$ branch. This kink is a further consequence of the vHS near $\ef$. The BCS result smears out the fine structure in $\langle n(\varepsilon_k)\rangle$ and is closer to the ideal non-interacting $T = 1\,$mK case than the strong-coupling Eliashberg calculation. 

\section{Conclusions}

We have demonstrated an efficient method to solve the Migdal-Eliashberg equations directly on the real-frequency axis at finite temperature while accounting for the electron-hole asymmetry introduced by the electronic density of states. Our linear-scaling numerical approach addresses the computational bottleneck of previous real-axis methods and avoids the need for analytic continuation from imaginary to real frequencies.

Applying these techniques to H$_3$S, we computed the temperature-dependent spectral function, quasiparticle density of states, and occupancies while capturing the particle-hole asymmetry induced by the van-Hove singularity near the Fermi level. The variable density of states treatment yields a low-temperature superconducting gap of ~60\,meV, matching experimental tunneling measurements and improving upon the 75\,meV predicted by the constant density of states approximation. This result illustrates the quantitative importance of retaining full electronic structure information in materials with strong particle-hole asymmetry.

The efficiency of our approach, with typical runtimes of milliseconds to minutes, combined with direct access to real-frequency Green's functions, facilitates the calculation of transport coefficients, optical conductivities, and time-dependent response functions. These capabilities make it practical to model nonequilibrium dynamics in superconducting devices and other phenomena that require real-frequency information. The methods outlined here provide a foundation for more systematic investigations of transport properties and nonequilibrium response in conventional superconductors within the framework of Migdal-Eliashberg theory.

\section{Computational Details}
DF(P)T calculations for H$_3$S were performed using the Quantum Espresso code \cite{QE-2009, QE-2017}, Perdew-Burke-Ernzerhof (PBE) exchange-correlation functional \cite{perdew_generalized_PBE_1996}, together with scalar-relativistic optimized norm-conserving Vanderbilt pseudopotentials \cite{vanbilt_pseudo_hamann_2013_PhysRevB.88.085117}. A $\mathbf{k}$-grid of $24\times24\times24$, a $\mathbf{q}$-grid of $4\times4\times4$, an energy cutoff of 100\,Ry, and a Methfessel-Paxton smearing \cite{Methfessel_PRB_1989_smearing} of 0.01\,Ry was applied. Using maximally localized Wannier functions, as implemented in the Wannier90 code \cite{Marzari2012, Pizzi2020}, and the EPW code \cite{PONCE2016116, Lee2023}, electron-phonon coupling matrix elements were computed on fine $48\times48\times48$ $\mathbf{k}$ and $\mathbf{q}$-grids. BerkeleyGW \cite{PhysRevB.34.5390, PhysRevB.62.4927, DESLIPPE20121269} was used to compute $W(\varepsilon, \varepsilon')$, with a dielectric energy cutoff of 25\,Ry and a $6\times6\times6$ $\mathbf{q}$-grid as reported in \cite{isoME}.

Further details on calculations can be found in Ref. \cite{lucrezi_full-bandwidth_2024} and \cite{isoME}.

\section{Acknowledgments}

This work was funded in part by the Defense Sciences Office (DSO) of the Defense Advanced Research Projects Agency (DARPA) (HR0011-24-9-0311). AS acknowledges support from the NSF GRFP. RF acknowledges support from the Alan McWhorter fellowship. PNF acknowledges support from the Austrian Science Fund (FWF) under project DOI 10.55776/ESP8588124. EK, PNF, MS, and CH acknowledge support from the Enterprise Science Fund of Intellectual Ventures and usage of computational resources of the lCluster of the Graz University of Technology and of the Austrian Scientific Computing (ASC) infrastructure.

\onecolumngrid

\section*{Supplemental Information}

\section{Derivation of Migdal-Eliashberg Equations}

We derive the isotropic Migdal-Eliashberg equations on the real axis within the vDOS+$W$ (variable electron density of states and static coulomb potential) approximation. The corresponding equations on the Matsubara axis are given in Ref.~\cite{isoME}.

The electronic self-energy on the Matsubara axis for a superconducting system in Nambu space  is given by~\cite{ALLEN19831}:
\begin{equation}
\label{eq:SelfEnergyGeneral}
    \Sigma(k,i\wj) = -\frac{1}{\beta}\sum_{k',j'}\tau_3 G(k,i\wjp) \tau_3 \left[W_{k,k'}(i\wj-i\wjp) + \sum_\lambda |g_{kk'\lambda}|^2 D_\lambda(k-k',i\wj-i\wjp)\right]~,
\end{equation}
where $\beta=\frac{1}{k_B T}$ is the inverse temperature, $k=(\bk,n)$ denotes the combined momentum and band index and $\lambda$ the phonon mode index.
The electron and phonon Green's functions can be expressed via their spectral representations:
\begin{equation}
    G(k,i\wj) = \int_{-\infty}^\infty d\omega \frac{A(k,\omega)}{i\wj-\omega}~,
\end{equation}
\begin{equation}
    D_{\lambda}(k-k',i\wj) = \int_{-\infty}^\infty d\Omega \frac{B_\lambda(k-k',\Omega)}{i\wj-\Omega}.
\end{equation}
Assuming a static Coulomb interaction and inserting the spectral representation of the phonon Green’s function, Eq.~\ref{eq:SelfEnergyGeneral} can be rewritten as:
\begin{equation}
\begin{split}
    \Sigma(k,i\wj) = -\frac{1}{\beta}\sum_{k',j'}\tau_3 G(k,i\wjp) \tau_3 \left[W_{k,k'} + \sum_\lambda |g_{kk'\lambda}|^2 \int_{-\infty}^\infty d\Omega \frac{B_\lambda(k-k',\Omega)}{i\wj-i\wjp-\Omega}\right] \\
    = -\frac{1}{\beta}\sum_{k',j'}\tau_3 G(k,i\wjp) \tau_3 \left[W_{k,k'} + \frac{1}{\NF} \int_{-\infty}^\infty d\Omega \frac{\afkko}{i\wj-i\wjp-\Omega} \right]~,
\end{split}
\end{equation}
where the Eliashberg spectral function has been introduced:
\begin{equation}
    \afkko = \NF \sum_\lambda  |g_{kk'\lambda}|^2 B_\lambda(k-k',\Omega)~.
\end{equation}
Next, inserting the spectral representation of the electronic Green’s function yields:
\begin{equation}
    \Sigma(k,i\wj) = -\frac{1}{\beta}\sum_{k',j'} \int d\omega' \frac{A(k',\omega')}{i\wjp-\omega'} \left[W_{k,k'} + \frac{1}{\NF} \int_{-\infty}^\infty d\Omega \frac{\afkko}{i\wj-i\wjp-\Omega} \right]~,
\end{equation}
This formally defines the analytic continuation to the real axis through the relation $i\wj\mapsto\omega+i0^+$.
The Pauli matrices have been added to the definition of the electronic spectral function:
\begin{equation}
\label{eq:elSpectralFunction}
    A(k,\omega) = -\frac{1}{\pi}\text{Im}\{\tau_3 G(k,\omega+i0^+)\tau_3\}~.
\end{equation}
For clarity, we separate the self-energy into an electron-phonon and Coulomb part:
\begin{equation}
    \Sigma_{ep} = \frac{1}{\beta\NF}\sum_{k'} \int d\omega' A(k',\omega') \int_{0}^\infty d\Omega\ \afkko \sum_{\nu} \frac{1}{i\wj-i\omega_\nu-\omega'}\frac{2\Omega}{(\omega_\nu)^2+\Omega^2}
\end{equation}
\begin{equation}
    \Sigma_C = -\frac{1}{\beta}\sum_{k'}  \int_{-\infty}^{\infty} d\omega' A^{\text{od}}(k',\omega') W_{k,k'} \sum_{j'} \frac{1}{i\wjp-\omega} 
\end{equation}
In the electron-phonon contribution, the summation over Matsubara frequencies was replaced by $i\omega_\nu=i\wj-i\wjp$ and the integration over $\Omega$ was restricted to positive values through the symmetry relation $\alpha^2F(-\Omega)=-\afo$.
The Coulomb contribution to the self-energy contains only the off-diagonal elements of $G^{\text{od}}(k,i\wj)$, as the Coulomb interaction is already contained in the band structure of the normal state. This constitutes an excellent approximation for the exact expression $G-G^N$, where $G^N$ is the Green's function describing the normal state. A detailed discussion can be found in chapter 9 of Ref.~\cite{ALLEN19831}.

The Matsubara sums can be evaluated via the identities~\cite{mahan}\footnote{Ref.~\cite{ALLEN19831} uses $\frac{1}{\beta}\sum_j \frac{1}{\xi-i\wj} = \frac{1}{2} \tanh{\frac{\beta \omega'}{2}}$, which is equivalent}:
\begin{equation}
    \frac{1}{\beta}\sum_j \frac{1}{i\wj-\xi} = \frac{1}{2}\big[2\nf(\xi)-1\big]
\end{equation}
and~\cite{ALLEN19831}:
\begin{equation}
    \frac{1}{\beta}\sum_\nu \frac{1}{i\wj-i\omega_\nu-\omega'}\frac{2\Omega}{\omega_\nu^2+\Omega^2} = \frac{\nb(\Omega)+1-\nf(\omega')}{i\wj-\Omega-\omega'}+\frac{\nb(\Omega)+\nf(\omega')}{i\wj+\Omega-\omega'} = I(i\wj,\Omega, \omega')~,
\end{equation}
where $\nf(\xi)$ and $\nb(\xi)$ denote the Fermi-Dirac and Bose-Einstein distribution, respectively.
Incorporating these identities and performing the analytic continuation leads to the anisotropic Eliashberg self-energy on the real axis:
\begin{equation}
    \Sigma_{ep} = \frac{1}{\NF}\sum_{k'} \int d\omega' A(k',\omega') \int_{0}^\infty d\Omega\ \afkko  \left[\frac{\nb(\Omega)+1-\nf(\omega')}{\omega-\Omega-\omega'+i0^+}+\frac{\nb(\Omega)+\nf(\omega')}{\omega+\Omega-\omega'+i0^+}\right]
\end{equation}
\begin{equation}
    \Sigma_C = -\frac{1}{2}\sum_{k'}  \int_{-\infty}^{\infty} d\omega' A^{\text{od}}(k',\omega') W_{k,k'} [2\nf(\omega')-1] 
\end{equation}
Finally, within the isotropic approximation the Eliashberg spectral function is averaged over the Fermi surface, while the Coulomb interaction is averaged over constant-energy surfaces~\cite{isoME, ALLEN19831}:
\begin{equation}
\begin{split}
    \afkko \mapsto \afo \\
    W_{k,k'} \mapsto W(\varepsilon,\varepsilon')
\end{split}
\end{equation}
and the momentum sums are replaced by energy integrals,
\begin{equation}
    \sum_k \mapsto \int_{-\infty}^{\infty} d\varepsilon N(\varepsilon)~,
\end{equation}
which yields the isotropic form of the self-energy:
\begin{equation}
\begin{split}
\label{eq:selfEnergyFinal}
        \Sigma(\varepsilon, \omega) &= -\frac{1}{\pi} \int_{-\infty}^{\infty} d\omega'  \int_{-\infty}^{\infty} d\varepsilon' N(\varepsilon') \Bigg\{ \text{Im}[\tau_3G(\varepsilon',\omega')\tau_3] \int_0^\infty d\Omega\ \frac{\afo}{\NF}\\ &\times\bigg[\frac{\nb(\Omega)+1-\nf(\omega')}{\omega-\Omega-\omega'+i0^+}+\frac{\nb(\Omega)+\nf(\omega')}{\omega+\Omega-\omega'+i0^+} \bigg] 
        - \frac{1}{2} \text{Im}[\tau_3G^{\text{od}}(\varepsilon',\omega')\tau_3]W(\varepsilon,\varepsilon')\big[2\nf(\omega')-1\big] \Bigg\}
\end{split}
\end{equation}
where the definition of the electronic spectral function from Eq.~\ref{eq:elSpectralFunction} has been inserted.

Another form of the self-energy is given by the usual decomposition using Pauli-matrices $\tau_i$:
\begin{equation}
\begin{split}
\label{eq:SelfEnergyComponents}
    \Sigma(k,\omega) = [1-Z(k,\omega)]\omega \tau_0 +\phi(k,\omega)\tau_1 + \chi(k,\omega)\tau_3 \\
    \mapsto \Sigma(\varepsilon,\omega) = [1-Z(\omega)]\omega \tau_0 +\phi(\varepsilon,\omega)\tau_1 + \chi(\omega)\tau_3~,
\end{split}
\end{equation}
where an isotropic form was assumed and the energy dependence of $Z$ and $\chi$ has been neglected~\cite{ALLEN19831}.
Through the Dyson equation
\begin{equation}
    G^{-1}(k,\omega) = G_0^{-1}(k,\omega) - \Sigma(k,\omega)~,
\end{equation}
with the non-interacting Green's function given by:
\begin{equation}
    G_0(k,\omega) = \frac{1}{\omega-(\varepsilon_k-\mu_F)+i0^+}~,
\end{equation}
a decomposition of the interacting Green's function can be derived:
\begin{equation}
    G(k,\omega) = \frac{\omega Z(k,\omega)\tau_0 +[\varepsilon_k-\mu_F+\chi(k,\omega)]\tau_3 +\phi(k,\omega)\tau_1}{\Theta(k,\omega)}~.
\end{equation}
The denominator is defined as:
\begin{equation}
    \Theta(k,\omega) = \text{det}\ G^{-1}(k,\omega) = \omega^2 Z^2(k,\omega) - [\varepsilon_k-\mu_F+\chi(k,\omega)]^2 -\phi^2(k,\omega)~.
\end{equation}
Multiplying the Pauli-matrices onto both sides and taking the imaginary part gives:
\begin{equation}
    \text{Im}[\tau_3 G(k,\omega)\tau_3] = \text{Im}\Big[\frac{\omega Z(k,\omega)}{\Theta(k,\omega)}\Big]\tau_0  + \text{Im}\Big[\frac{\varepsilon_k-\mu_F+\chi(k,\omega)}{\Theta(k,\omega)}\Big]\tau_3 - \text{Im}\Big[\frac{\phi(k,\omega)}{\Theta(k,\omega)}\Big]\tau_1~.
\end{equation}
Within the isotropic approximation, this simplifies to
\begin{equation}
    \text{Im}[\tau_3 G(\omega)\tau_3] =\text{Im}\Big[\frac{\omega Z(\omega)}{\Theta(\varepsilon,\omega)}\Big]\tau_0  + \text{Im}\Big[\frac{\varepsilon-\mu_F+\chi(\omega)}{\Theta(\varepsilon,\omega)}\Big]\tau_3 - \text{Im}\Big[\frac{\phi(\varepsilon,\omega)}{\Theta(\varepsilon,\omega)}\Big]\tau_1~,
\end{equation}
with an isotropic denominator:
\begin{equation}
        \Theta(\varepsilon,\omega) =\omega^2 Z^2(\omega) - [\varepsilon-\mu_F+\chi(\omega)]^2 -\phi^2(\varepsilon,\omega)~.
\end{equation}
Inserting this expression into Eq.~\ref{eq:selfEnergyFinal} and comparing it to Eq.~\ref{eq:SelfEnergyComponents} results in a set of coupled equations - the real axis isotropic Eliashberg equations in the vDOS+$W$ approximation:
\begin{subequations}
    \begin{align}
    \begin{split}
        \label{eq:realvDOSW_Z}
        Z(\omega) = 1 + \frac{1}{\omega\pi\NF}&\int_{-\infty}^\infty d\omega' \int_{-\infty}^\infty d\varepsilon' N(\varepsilon') \text{Im}\Big[\frac{Z(\omega')\omega'}{\Theta(\varepsilon',\omega')}\Big] \\
        \times &\int_{0}^{\infty}d\Omega \afo \bigg[\frac{\nb(\Omega)+1-\nf(\omega')}{\omega-\Omega-\omega'+i0^+}+\frac{\nb(\Omega)+\nf(\omega')}{\omega+\Omega-\omega'+i0^+} \bigg]
    \end{split} \\
    \begin{split}
        \label{eq:realvDOSW_chi}
        \chi(\omega) = - \frac{1}{\pi\NF}&\int_{-\infty}^\infty d\omega' \int_{-\infty}^\infty d\varepsilon' N(\varepsilon') \text{Im}\Big[\frac{\varepsilon'-\mu_F+\chi(\omega')}{\Theta(\varepsilon',\omega')}\Big]\\
        \times &\int_{0}^{\infty}d\Omega \afo \bigg[\frac{\nb(\Omega)+1-\nf(\omega')}{\omega-\Omega-\omega'+i0^+}+\frac{\nb(\Omega)+\nf(\omega')}{\omega+\Omega-\omega'+i0^+} \bigg]
    \end{split}\\
    \begin{split}
        \label{eq:realvDOSW_phi}
        \phi(\varepsilon,\omega) = \frac{1}{\pi\NF}&\int_{-\infty}^\infty d\omega' \int_{-\infty}^\infty d\varepsilon' N(\varepsilon') \text{Im}\Big[\frac{\phi(\varepsilon',\omega')}{\Theta(\varepsilon',\omega')}\Big] \Bigg\{\int_{0}^{\infty}d\Omega \afo  \\
        \times&\bigg[\frac{\nb(\Omega)+1-\nf(\omega')}{\omega-\Omega-\omega'+i0^+}+\frac{\nb(\Omega)+\nf(\omega')}{\omega+\Omega-\omega'+i0^+} \bigg] - \frac{1}{2}\NF W(\varepsilon,\varepsilon') [2\nf(\omega')-1]\Bigg\}
    \end{split}
    \end{align}
\end{subequations}

Using the symmetry relations $Z(-\omega) = Z(\omega)^*$, $\phi(-\omega) = \phi(\omega)^*$, and $\chi(-\omega)=\chi(\omega)^*$ as derived in Section \ref{sec:symmG}, we can rewrite the equations with integrals over $\omega'$ only ranging from $0$ to $\infty$. 

\subsection{Electron number equation}
The electron number in the superconducting phase is given by (Eq.~(4) in the supplemental of Ref.~\cite{lucrezi_full-bandwidth_2024}):
\begin{equation}
    \Ne = 2 \sum_k \left\{\frac{1}{2} + k_B T \sum_j \text{Re} \left[ G_{11}(k,\omega) \right] \right\}
\end{equation}
where $G_{11}(k,\omega)$ is the first diagonal element of the matrix Green's function. Inserting the spectral representation yields:
\begin{equation}
\begin{split}
    \Ne &= 2 \sum_k \left\{ \frac{1}{2} + k_B T \sum_j \int_{-\infty}^{\infty}d\omega\ \text{Re} \left[ \frac{A_{11}(k,\omega)}{i\wj-\omega} \right] \right\} \\
    &=  2 \sum_k \left\{ \frac{1}{2} + k_B T \sum_j \int_{-\infty}^{\infty}d\omega\ \text{Re} \left[ \frac{A_{11}(k,\omega)\ (i\wj+\omega)}{(i\wj-\omega)(i\wj+\omega)} \right] \right\} \\
    &= 2 \sum_k \left\{ \frac{1}{2} + k_B T \sum_j \int_{-\infty}^{\infty}d\omega\ \frac{A_{11}(k,\omega)\ \omega}{(i\wj)^2-\omega^2} \right\} 
\end{split}        
\end{equation}
The Matsubara sum can be evaluated through contour integration~\cite{mahan}. For the given integrand, the contribution of the circle at infinity vanishes and we will deform the contour to half-circles $C_1$/$C_2$ in the negative ($z<0$) and positive ($z>0$) half-plane and thereby pick up only the contributions from the poles at $z=\pm \omega$.
\begin{equation}
    \frac{1}{\beta}\sum_j \frac{1}{(i\wj)^2-\omega^2} = \oint_{C_1}\frac{dz}{2\pi i} \frac{f(z)}{z^2-\omega^2} +\oint_{C_2}\frac{dz}{2\pi i} \frac{f(z)}{z^2-\omega^2}= \frac{1}{2\omega} [2\nf(\omega)-1]~,
\end{equation}
Inserting this result and the definition of the spectral function from Eq.~\ref{eq:elSpectralFunction} into the electron number equation gives:
 \begin{equation}
     \Ne=\sum_k \left\{ 1 - \frac{1}{\pi} \int_{-\infty}^{\infty}d\omega \, \text{Im} \left[ \frac{\varepsilon_k-\mu_F+\chi(k,\omega)}{\Theta(k,\omega)} \right] [2\nf(\omega)-1]\right\}~,
\end{equation}
which reduces to
\begin{equation}
         \Ne= \int d\varepsilon N(\varepsilon)\left\{ 1 - \frac{1}{\pi} \int_{-\infty}^{\infty} d\omega \,  \text{Im}\left[ \frac{\varepsilon-\mu_F+\chi(\omega)}{\Theta(\varepsilon,\omega)} \right] [2\nf(\omega)-1]\right\}
\end{equation}
when employing the isotropic approximation.

\section{Other Approximations}

\subsection{vDOS+$\mu$}

In the vDOS+$\mu$-approximation, we neglect the energy dependence of the static Coulomb interaction, i.e. replace $W(\varepsilon, \varepsilon')$ with $W(\ef, \ef)$. This removes the $\varepsilon$-dependence of $\phi(\varepsilon, \omega)$ so that $\phi(\varepsilon, \omega) = \phi(\omega)$. Introducing the abbreviation $\mu = \NF W(\ef, \ef)$ and restricting the $\omega'$ integration in Eq.~\ref{eq:realvDOSW_phi} to $|\omega'| \leq \wc$,
Eq.~\ref{eq:realvDOSW_phi} reduces to:
\begin{equation}
\label{eq:realvDOSmu_phi}
\begin{split}
    \phi(\omega) = \frac{1}{\pi\NF}&\int_{-\wc}^{\wc} d\omega' \int_{-\infty}^{\infty} d\varepsilon' N(\varepsilon') \text{Im}\Big[\frac{\phi(\varepsilon',\omega')}{\Theta(\varepsilon',\omega')}\Big] \Bigg\{\int_{0}^{\infty}d\Omega \afo  \\
    \times&\bigg[\frac{\nb(\Omega)+1-\nf(\omega')}{\omega-\Omega-\omega'+i0^+}+\frac{\nb(\Omega)+\nf(\omega')}{\omega+\Omega-\omega'+i0^+} \bigg] - \frac{1}{2}\muc(\omega_c) [2\nf(\omega')-1]\Bigg\}~,
\end{split} 
\end{equation}
where $\muc$ is the Morel-Anderson Pseudopotential ~\cite{morel_anderson_1962, ALLEN19831, ScalapinoSchrieffer}.
The other two equations retain their formal structure, with only the definition of $\Theta$ being modified to reflect the aforementioned changes:
\begin{equation}
    \Theta(\varepsilon,\omega)= \omega^2 Z^2(\omega) - [\varepsilon-\mu_F+\chi(\omega)]^2 -\phi^2(\omega)~.
\end{equation}

\subsection{cDOS+$\mu$}
In the cDOS+$\mu$ approximation, the electronic density of states is assumed to be constant. The only remaining $\varepsilon$-dependence is that of $\Theta(\varepsilon,\omega)$, which can be removed by integrating:
\begin{equation}
\label{eq:IntEpsiloncDOS}
    \int_{-\infty}^\infty d\varepsilon \frac{1}{\omega^2 Z^2(\omega) - \varepsilon^2 -\phi^2(\omega)}~.
\end{equation}
The integrand has poles at:
\begin{equation}
\begin{split}
    \varepsilon^2 &= -\phi^2(\omega)+\omega^2 Z^2(\omega) \\
    \Rightarrow \varepsilon &= \pm i \sqrt{-\omega^2Z^2(\omega)+\phi^2(\omega)}=\pm i \sqrt{C} ~. \\
    C &= \phi^2(\omega) -\omega^2Z^2(\omega) 
\end{split}
\end{equation}
By closing the contour in the upper half plane, we obtain:
\begin{equation}
   -\oint dz \frac{1}{z^2+C} = -2\pi i \text{Res}\left[\frac{1}{z^2+C}\right]_{z=+ i\sqrt{C}}=-2\pi i \frac{1}{2i\sqrt{C}}= -\frac{\pi}{\sqrt{C}}~~.
\end{equation}
Inserting this into Eq.~\ref{eq:IntEpsiloncDOS} yields:
\begin{equation}
    \int_{-\infty}^\infty d\varepsilon \frac{1}{\omega^2 Z^2(\omega) - \varepsilon^2 -\phi^2(\omega)} = -\frac{\pi}{\sqrt{- (\omega^2Z^2(\omega)-\phi^2)}}= \frac{i\pi}{\sqrt{\omega^2Z^2(\omega)-\phi^2}}
\end{equation}
and the equations in the cDOS+$\mu$ approximation are simplified to:
\begin{subequations}
    \begin{align}
    \begin{split}
    \label{eq:realcDOSmu_Z}
        Z(\omega) = 1 - \frac{1}{\omega} &\int_{-\infty}^\infty d\omega'\text{Re}\left[\frac{\omega' Z(\omega')}{\sqrt{\omega'^2Z^2(\omega') -\phi^2(\omega')}}\right]  \\
        \times  & \int_0^\infty d\Omega \afo\bigg[\frac{\nb(\Omega)+1-\nf(\omega')}{\omega-\Omega-\omega'+i0^+}+\frac{\nb(\Omega)+\nf(\omega')}{\omega+\Omega-\omega'+i0^+} \bigg]
    \end{split} \\
    \begin{split}
    \label{eq:realcDOSmu_phi}
        \phi(\omega) = -&\int_{-\wc}^{\wc} d\omega'\text{Re}\left[\frac{\phi(\omega')}{\sqrt{\omega'^2Z^2(\omega') -\phi^2(\omega')}}\right]  \Bigg\{\int_0^\infty d\Omega \afo  \\
         \times &\bigg[\frac{\nb(\Omega)+1-\nf(\omega')}{\omega-\Omega-\omega'+i0^+}+\frac{\nb(\Omega)+\nf(\omega')}{\omega+\Omega-\omega'+i0^+} \bigg] -\frac{1}{2}\muc(\omega_c)[2\nf(\omega')-1] \Bigg\}~,
    \end{split}
    \end{align}
\end{subequations}
where the identity $\text{Im}[iz] = \text{Re}[z]$ was used. In analogy to Eq.~(64) of Ref.~\cite{parks2018superconductivity} (part 10 of Scalapino, page 496) and Eq.~(12.8) of Ref.~\cite{ALLEN19831} the square root is defined to have a positive imaginary part when $\omega$ is in the upper half-plane. 

\section{Symmetries of $G$}
\label{sec:symmG}
In Nambu space, the following relation holds
\begin{equation}
    G^R(\omega) = -\tau_2 \left[ G^R(-\omega) \right]^* \tau_2 \label{eq:GRsymm}
\end{equation}
Recalling the form
\begin{align}
    G^R(\omega) =& 
    \begin{pmatrix}
\omega Z(\omega) + \chi(\omega) & \phi(\omega) \\
\phi(\omega) & \omega Z(\omega) - \chi(\omega) 
\end{pmatrix}
\end{align}
we obtain
\begin{align}
-\tau_2 \left[ G^R(-\omega) \right]^* \tau_2 =& 
\begin{pmatrix}
\omega Z^*(-\omega) + \chi^*(-\omega) & \phi^*(-\omega) \\
\phi^*(-\omega) & \omega Z^*(-\omega) - \chi^*(-\omega) 
\end{pmatrix}
\end{align}
Equating matrix entries yields
\begin{subequations}
\begin{align}
\omega Z(\omega) + \chi(\omega) =& \omega Z^*(-\omega) + \chi^*(-\omega) \label{eq:Zchi_symm1} \\
\omega Z(\omega) - \chi(\omega) =& \omega Z^*(-\omega) - \chi^*(-\omega) \label{eq:Zchi_symm2} \\
\phi(\omega) =& \phi^*(-\omega) \label{eq:phi_symm}
\end{align}
\end{subequations}

Adding Eq.~\ref{eq:Zchi_symm1} and Eq.~\ref{eq:Zchi_symm1} gives
\begin{equation}
    Z(\omega) = Z^*(-\omega) \, \label{eq:Z_symm}
\end{equation}
and subtracting them gives
\begin{equation}
    \chi(\omega) = \chi^*(-\omega) \, .\label{eq:chi_symm}
\end{equation}
Thus the real parts of $Z, \chi$, and $\phi$ are even functions for real-valued $\omega$ and their imaginary parts are of odd parity. Using this, once can rewrite the integrals over omega from a symmetric form, i.e., $\int \limits_{-\infty}^\infty d\omega ...$ to only span the positive frequency region $\int \limits_{0}^\infty  d\omega ...$. In practice, these symmetries reduce the required numerical computation by half.

A similar expression for the symmetry of $G$ can also be derived from Lehmann's representation of $G$, see for example Ref.~\cite{khodachenko2024nevanlinna} (paragraph above Eq.~(21))
\begin{subequations}
\begin{align}
    G_{11}(\omega) =& \sum_\alpha \frac{|\langle \alpha | \hat{c}| 0\rangle|^2}{\omega - (E_\alpha - E_0) + i\delta} +  \sum_\beta \frac{|\langle \beta | \hat{c}^\dagger| 0\rangle|^2}{\omega + (E_\beta - E_0) + i\delta} \label{eq:LG11} \\
    G_{22}(\omega) =& \sum_\beta \frac{|\langle \beta | \hat{c}^\dagger| 0\rangle|^2}{\omega - (E_\beta - E_0) + i\delta}  + \sum_\alpha \frac{|\langle \alpha | \hat{c}| 0\rangle|^2}{\omega + (E_\alpha - E_0) + i\delta} \label{eq:LG22} \, .
\end{align}
\end{subequations}
In order to match the particle ($\hat{c}^\dagger$) and the hole ($\hat{c}$) terms upon a transformation $\omega \rightarrow -\omega$, one has to multiply the respective terms with a factor of $(-1)$ and take the complex conjugate to regain the correct sign of the $i\delta$ term.

This means in short that
\begin{equation}
    G_{11}(-\omega) = - \left[ G_{22}(\omega)\right]^* \, ,
\end{equation}
consistent with Eq.~\ref{eq:GRsymm}. For the anomalous parts one has the usual
\begin{equation}
    F_{12}(-\omega) =  F^*_{21}(\omega) \, .
\end{equation}

\section{Analytic continuation with the Padé approximation}
We compare the results of our real-frequency axis solver against the analytically continued solution on the imaginary frequency axis obtained using the isoME package \cite{isoME}. Here, we employ the Padé approximation to perform the analytic continuation; however, as discussed in the main text, there are several possible approaches that one can adopt here \cite{baker1961pade, kraberger2017maximum, vitali2010ab, PhysRevB.61.5147, khodachenko2024nevanlinna}. 

\begin{figure*}
    \centering
    \includegraphics[width=\linewidth]{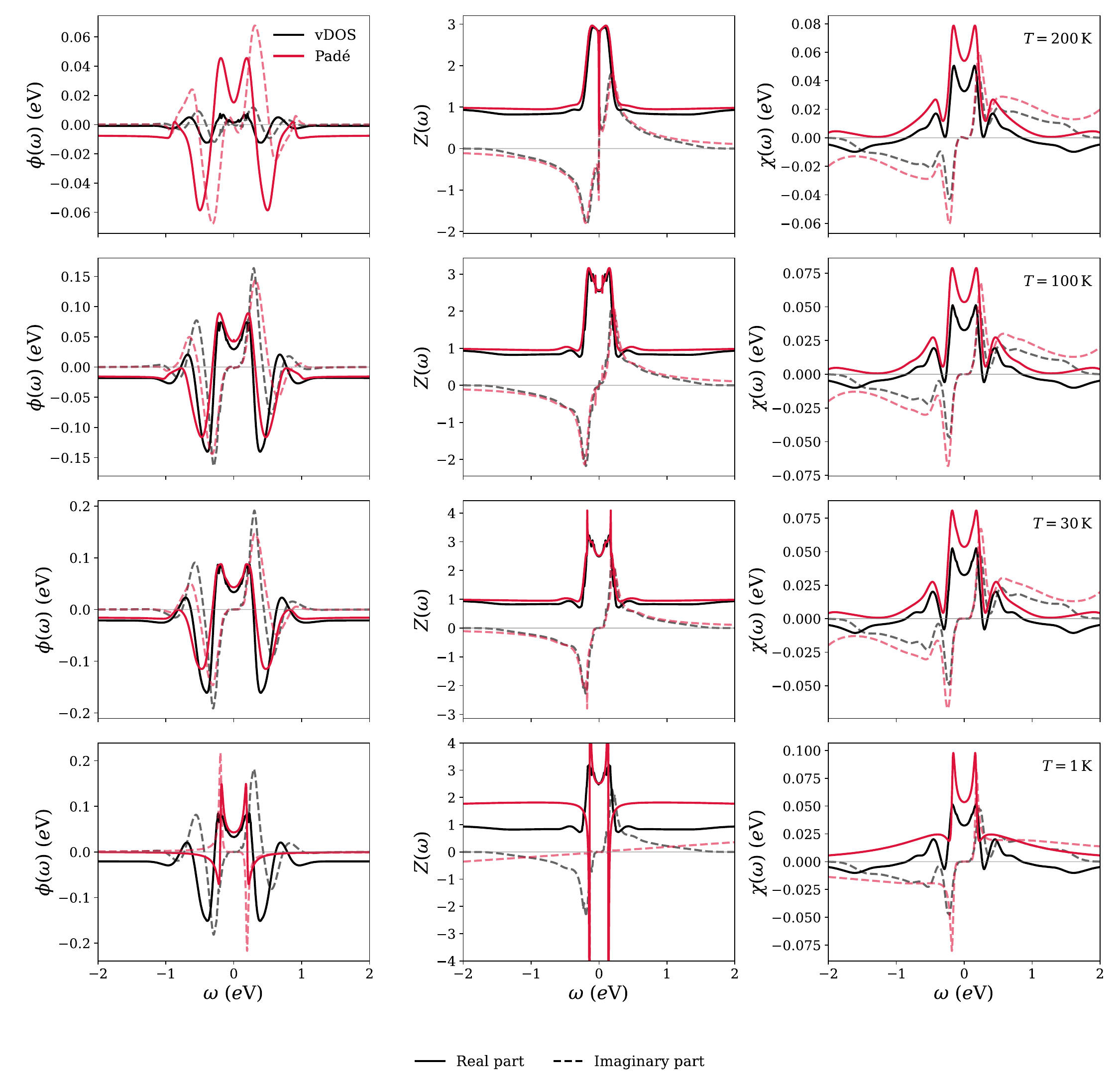}
    \caption{Comparison between the direct real-frequency-axis vDOS solutions (black) and the analytic continuation (red) done from the imaginary frequency axis with the Padé approximation for $\phi(\omega)$, $Z(\omega)$, and $\chi(\omega)$ of H$_3$S at temperatures $T=200\,$K, $100\,$K, $30\,$K, and $1\,$K. The real part of the solutions is shown in solid lines and the imaginary part corresponds to the dashed lines. Numerical instabilities in $Z(\omega)$ are clear in the analytically continued solutions.}
    \label{fig:Pade}
\end{figure*}

In Figure \ref{fig:Pade}, the analytically continued result for H$_3$S is shown alongside the direct real-frequency solution. There is good qualitative agreement between $\phi(\omega)$, $Z(\omega)$, and $\chi(\omega)$ obtained with the two methods, indicating that the direct real-frequency axis solutions are properly converging. As expected, the fine structure in the solutions for these quantities is obscured in the analytically continued result. This is particularly evident in the absence of peaks in $\phi(\omega)$ corresponding with the structure of $\alpha^2F(\omega)$. With the direct solutions, the numerical stability is also improved over a larger range of temperatures. Instabilities in the continuation procedure are particularly evident in $Z(\omega)$ as the temperature range is swept from $T= 1\,$K to 200$\,$K, manifesting as large unphysical peaks in the magnitude of $Z(\omega)$. This is to be contrasted with the smooth solutions obtained with the direct vDOS real-frequency solutions. Finally, the absolute magnitude of $\chi(\omega)$ depends on the cutoff chosen when evaluating the Migdal-Eliashberg equations. Hence, $\chi(\omega)$ is not expected to quantitatively agree between two sets of solutions obtained with different parameters, which explains the quantitative disagreement in Figure \ref{fig:Pade}. The strong qualitative agreement is expected.

\section{Numerical Methods}

\subsection{Evaluation of Kernel Integrals}

\begin{figure}
    \centering
    \includegraphics[width=\linewidth]{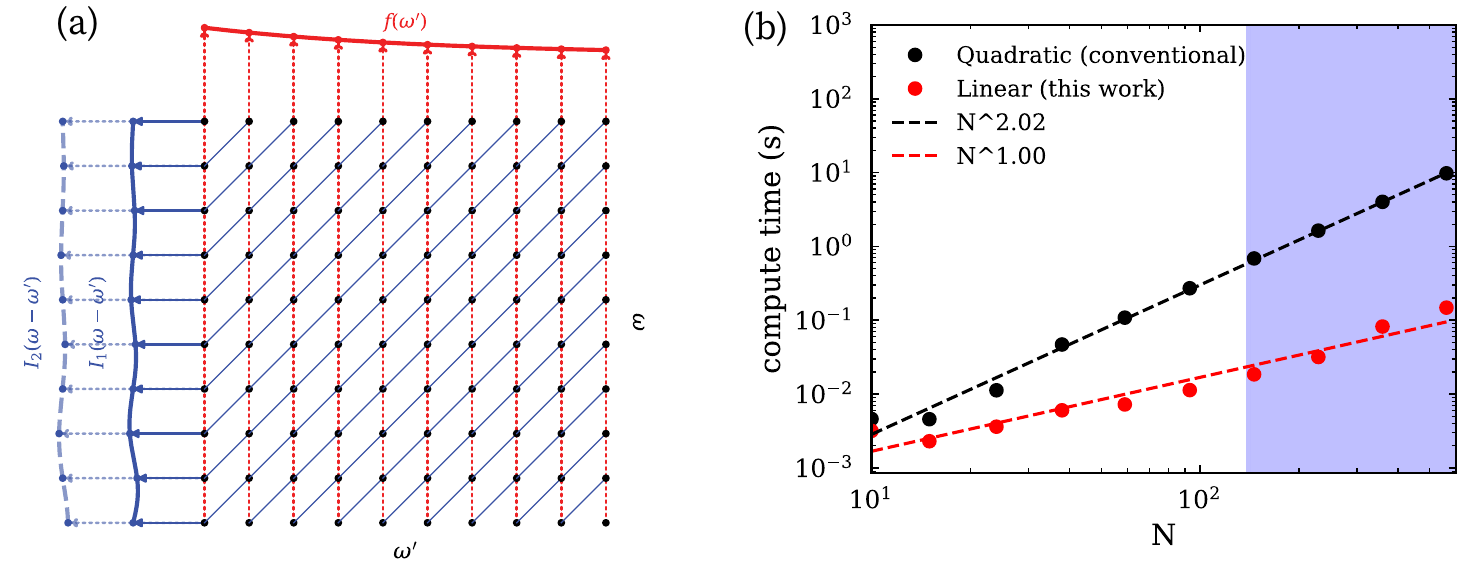}
    \caption{(a) Schematic for linear-time computation of $K(\omega, \omega')$ over a square grid of points. Blue diagonals pass through all points sharing a common value of $\omega-\omega'$, and vertical lines pass through all points with the same $\omega'$. The integrals $I_1$ and $I_2$ need only be evaluated once and reused by all points in each blue diagonal. (b) Runtime comparison for our numerical approach to compute $K(\omega,\omega')$ versus the standard quadratic computation of the kernels. The highlighted blue region indicates $N$ large enough for the solution to converge.}
    \label{fig:speed}
\end{figure}

The evaluation of the kernel integrals has traditionally been a computational bottleneck for direct solutions to the Eliashberg equations on the real axis. We wish to compute 
\begin{equation}
    K(\omega, \omega') = \int_{0}^{\infty} d\Omega \afo \bigg[\frac{\nb(\Omega)+1-\nf(\omega')}{\omega-\Omega-\omega'+i0^+}+\frac{\nb(\Omega)+\nf(\omega')}{\omega+\Omega-\omega'+i0^+} \bigg]
\end{equation} on an evenly spaced square grid of $(\omega, \omega')$ points. Suppose that we take $N$ points along each axis. A naive approach would require $O(N^2)$ integral evaluations. We will demonstrate an efficient algorithm that requires only $O(N)$ evaluations. The imaginary parts $\text{Im} \{K(\omega, \omega')\}$ can be quickly evaluated because they amount to integrating over a delta function. For the real parts, we can write
\begin{equation}
\label{eq:reduction}
    \text{Re}\{K(\omega, \omega')\} = [-I_1(\omega -\omega') - (1-f(\omega'))I_2(\omega-\omega')] + [I_1(\omega'-\omega) + f(\omega')I_2(\omega'-\omega)]
\end{equation}
where
\begin{equation}
\begin{aligned}
    &I_1(x) = \mathcal P \int_{-\infty}^\infty d\Omega \, \alpha^2F(\Omega) \frac{n(\Omega)}{\Omega - x}\\
    &I_2(x) = \mathcal P \int_{-\infty}^\infty d\Omega \, \alpha^2F(\Omega) \frac{1}{\Omega - x}
\end{aligned}
\end{equation}
On a square grid of $(\omega, \omega')$ points, there are only $2N-1$ distinct values of $\omega-\omega'$, as illustrated in Figure \ref{fig:speed}a. Thus, by evaluating $I_1$ and $I_2$ for all such values, we can produce the full matrix of kernel values by computing only $O(N)$ of the costly principal value integrals. Materializing the full matrix and multiplying in the factors of $f(\omega')$ presents a slight overhead, but in practice this algorithm allows us to precompute $K(\omega, \omega')$ in less than a second.

To actually evaluate the principal value integrals $I_1$ and $I_2$ on a grid of $s$ values, we must account for the divergence of the integrands at $\Omega=s$. We employ a change of variables $\Omega' = \Omega - s$ to shift all singularities to $\Omega'=0$. This allows us to use a common grid of $\Omega'$ samples for the entire $s$ grid and evaluate the integrals in a vectorized manner. We employ a symmetric grid of Chebyshev nodes around the singularity in a Clenshaw–Curtis quadrature scheme.

We show in Figure \ref{fig:speed}b a benchmark comparing our numerical approach to the standard quadratic-time evaluation of the integral, where the runtime is plotted as a function of the number of sampling points $N$. In the conventional implementation, the cost scales as $O(N^2)$, since $K(\omega,\omega')$ must be evaluated for all pairs of sampling points. In contrast, evaluating $K(\omega,\omega')$ using Eq. \eqref{eq:reduction} results in a much more favorable linear $O(N)$ cost. The exponential fit in Figure \ref{fig:speed}b highlights this rapid reduction in computational costs. This improvement enables well-converged and efficient evaluations of the integral in regimes where the quadratic approach becomes prohibitively expensive.

\subsection{Evaluation of Electronic Spectral Function Integrals}

We must evaluate integrals of the form
\begin{equation}
    I_\star[g(\varepsilon, \omega)] = \int_{-\infty}^\infty d\varepsilon\, N(\varepsilon) \Im \left[\frac{g(\varepsilon, \omega)}{\Theta(\varepsilon, \omega)} \right]
\end{equation}
to high precision. In the cDOS approximation, the integrals are easy to evaluate analytically using the residue theorem. However, when $N(\varepsilon)$ is non-constant, the integrals become numerically challenging due to sharp peaks near the poles of the Green's function. To avoid these challenges, we leverage the form of $\text{Im} \left[\frac{g(\varepsilon, \omega)}{\Theta(\varepsilon, \omega)} \right]$ to derive a semi-analytic quadrature method. 

In the vDOS+$\mu$ case, $\phi(\varepsilon, \omega) = \phi(\omega)$ is independent of $\varepsilon$. We can do a partial fraction decomposition:
\begin{equation}
    \Im \left[\frac{g(\varepsilon, \omega)}{\Theta(\varepsilon, \omega)} \right] = \Im \left[\frac{g(\varepsilon, \omega) / 2\varepsilon_p(\omega)}{\varepsilon + \chi(\omega) + \varepsilon_p(\omega)} \right] - \Im \left[\frac{g(\varepsilon, \omega) / 2\varepsilon_p(\omega)}{\varepsilon + \chi(\omega) - \varepsilon_p(\omega)} \right]
\end{equation}
where $\varepsilon_p = \sqrt{\omega^2 Z(\omega)^2 - \phi(\omega)^2}$. Define $R_\pm(\omega) = \Re\{\chi(\omega) \pm \varepsilon_p(\omega)\}$ and $I_\pm(\omega) = \Im\{\chi(\omega) \pm \varepsilon_p(\omega)\}$. Also define $R_g = \Re \{ g(\varepsilon, \omega)/2\varepsilon_p\}$ and $I_g = \Im \{ g(\varepsilon, \omega)/2\varepsilon_p\}$. We can express each partial fraction above with real coefficients:
\begin{equation}
    \Im \left[ \frac{g(\varepsilon, \omega) / 2\varepsilon_p(\omega)}{\varepsilon + \chi(\omega) \pm \varepsilon_p(\omega)}\right] = I_g \frac{\varepsilon}{(\varepsilon+R_\pm)^2 + I_\pm^2} + (R_\pm I_g - I_\pm R_g) \frac{1}{(\varepsilon+R_\pm)^2 + I_\pm^2}.
\end{equation}
In the vDOS+$\mu$ case, $g(\varepsilon,\omega)$ has either constant or linear dependence on $\varepsilon$. Thus, within a given interval where $N(\varepsilon)$ is approximated as linear, the integrand in (6) can be written as a sum of terms, each of which is a polynomial of degree at most $3$ multiplying a Lorentzian. Note that there exists an analytic antiderivative of a Lorentzian function and any polynomial multiple of a Lorentzian.

\begin{figure}
    \centering
    \includegraphics[width=0.5\linewidth]{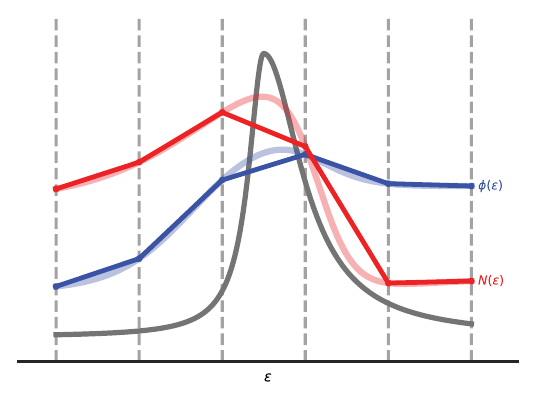}
    \caption{Illustration of spectral function integration. The functions $\phi(\varepsilon, \omega)$ and $N(\varepsilon)$ are approximated as piecewise linear functions. The gray curve shows the Lorentzian-like function of $\varepsilon$ that multiplies $N(\varepsilon)$.}
    \label{fig:sampling}
\end{figure}

This motivates us to interpolate a piecewise-linear approximation for $N(\varepsilon)$ as depicted in Figure \ref{fig:sampling}, so that within each interval we can analytically integrate the approximation over the expression derived above. More formally, suppose that we have a sequence of evenly spaced data points $\{(\varepsilon_j, N_j)\}_{j=1}^m$, ordered by $\varepsilon$ values. Define $\tilde N_m(\varepsilon)$ on the interval $[\varepsilon_1, \varepsilon_m]$ such that if $\varepsilon \in [\varepsilon_j, \varepsilon_{j+1}]$, then $\tilde N(\varepsilon) = \frac{N_{j+1} - N_j}{\varepsilon_{j+1} - \varepsilon_j}(\varepsilon - \varepsilon_j) + N_j = M^{(j)}_0 + M^{(j)}_1 \varepsilon$. Then we compute
\begin{equation}
    \int_{-\infty}^\infty d\varepsilon\, \tilde N(\varepsilon) \Im \left[\frac{g(\varepsilon, \omega)}{\Theta(\varepsilon, \omega)} \right] = \sum_{j=1}^{m-1} \int_{\varepsilon_j}^{\varepsilon_{j+1}} (M_0^{(j)} + M_1^{(j)} \varepsilon) \, \Im \left[\frac{g(\varepsilon, \omega)}{\Theta(\varepsilon, \omega)} \right]
\end{equation}
Note that the approximation approaches the actual integral value in the limit of infinitely many sample points, as $\lim_{m \to \infty} \tilde N_m(\varepsilon) = N(\varepsilon)$.

The vDOS+$W$ case can be transformed into the vDOS+$\mu$ case by similarly using a piecewise linear approximation along the $\varepsilon$ axis for $\varphi(\varepsilon, \omega)$. Approximate $\phi(\varepsilon, \omega) \approx \tilde \phi(\varepsilon, \omega) = \Phi_0^{(j)}(\omega) + \Phi_1^{(j)}(\omega)\varepsilon$ on each interval $[\varepsilon_j, \varepsilon_{j+1}]$. Then for $\mu_F = 0$ we have
\begin{equation}
    \begin{aligned}
        \Theta(\varepsilon, \omega) &\approx \omega^2 Z(\omega)^2 - (\varepsilon + \chi(\omega))^2 - (\Phi_0^{(j)}(\omega) + \Phi_1^{(j)}(\omega)\varepsilon)^2\\
        &= \omega^2 Z(\omega)^2 - \varepsilon^2 - 2\varepsilon\chi(\omega) - \chi(\omega)^2 - \Phi_0^{(j)}(\omega)^2 - 2\Phi_0^{(j)}(\omega)\Phi_1^{(j)}(\omega) - \Phi_1^{(j)}(\omega)^2 \varepsilon^2\\
        &= \omega^2 Z(\omega)^2 - (1+\Phi_1^{(j)}(\omega)^2) \left[ \varepsilon^2 + \frac{2\varepsilon(\chi(\omega) + \Phi_0^{(j)}(\omega)\Phi_1^{(j)}(\omega))}{1 + \Phi_1^{(j)}(\omega)^2}\right] - \chi(\omega)^2 - \Phi_0^{(j)}(\omega)^2\\
        &= (1+\Phi_1^{(j)}(\omega)^2) \left[ \frac{\omega^2 Z(\omega)^2 - \chi(\omega)^2 - \Phi_0^{(j)}(\omega)^2}{1 + \Phi_1^{(j)}(\omega)^2} + S^{(j)}(\omega)^2 - (\varepsilon + S^{(j)}(\omega))^2\right]
    \end{aligned}
\end{equation}
where $S^{(j)}(\omega) = \frac{\chi(\omega) + \Phi_0^{(j)}(\omega)\Phi_1^{(j)}(\omega)}{1 + \Phi_1^{(j)}(\omega)^2}$. Define $P^{(j)}(\omega)^2 = \frac{\omega^2 Z(\omega)^2 - \chi(\omega)^2 - \Phi_0^{(j)}(\omega)^2}{1 + \Phi_1^{(j)}(\omega)^2} + S^{(j)}(\omega)^2$. Then we can write
\begin{equation}
    \Theta(\varepsilon, \omega) \approx (1 + \Phi_1^{(j)}(\omega)^2)(P^{(j)}(\omega)^2 - (\varepsilon + S^{(j)}(\omega))^2)
\end{equation}
so that $P^{(j)}(\omega)$ and $S^{(j)}(\omega)$ are analogous to $\varepsilon_p(\omega)$ and $\chi(\omega)$ in the formulas derived for the vDOS+$\mu$ case. We can also absorb the $1 + \Phi_1^{(j)}(\omega)^2$ factor into $g(\varepsilon, \omega)$.

\subsection{Causality of the Green's Function}

The poles of the Green's function, which give the quasiparticle excitations, are the roots of $\Theta(\varepsilon, \omega)$. From causality, we require  the imaginary part of each pole to have the same sign as the real part for all $\omega$. The converged solution to the Eliashberg equations must obey this condition, but intermediate guesses in the iteration step may not. This point is crucial, for violation of this constraint amounts to an overall sign flip in the spectral function integrals. Even if this occurs for only a small fraction of $\omega$ values, it creates jump discontinuities in the spectral function integrals as a function of $\omega$. We observed that such discontinuities cause the fixed-point iterations to diverge.

In order to remedy this issue, we manually enforce a sign flip on the spectral function integrals when the imaginary parts of the Green's function poles have the wrong sign. For the vDOS+$\mu$ case, this is easy, as we can directly compute the poles. However, in the vDOS+$W$ case, only knowing $\phi(\varepsilon, \omega)$ on the real $\varepsilon$ axis means we cannot determine the complex poles of the Green's function. To circumvent this obstacle, we notice that the spectral function integral in the equation for $Z(\omega)$ is the quasiparticle density of states, which must always be positive. Thus, the sign flip is equivalent to taking the absolute value of the spectral function integral in this case. We can identify which values were flipped by this operation and, since the sign flip must be applied to the same $\omega$ values across all three integrals, this allows us to correctly enforce the sign flip across all three integrals.



%

\end{document}